\documentclass[preprints,article,accept,moreauthors,pdftex]{Definitions/mdpi} 
\usepackage{enumitem}
\usepackage{upgreek}

\firstpage{1} 
\pubvolume{1}
\issuenum{1}
\articlenumber{0}
\pubyear{2021}
\copyrightyear{2021}
\externaleditor{Academic Editor: Jose L. Gómez} 
\datereceived{18 October 2021} 
\dateaccepted{6 December 2021} 
\datepublished{} 
\hreflink{https://doi.org/} 

\newcommand\aj{AJ}

\newcommand\araa{ARA\&A}
\newcommand\apj{ApJ}
\newcommand\apjl{ApJ}

\newcommand\aap{A\&A}
\newcommand\aapr{A\&A~Rev.}
\newcommand\aaps{A\&AS}

\newcommand\mnras{MNRAS}

\newcommand\pasa{PASA}
\newcommand\pasp{PASP}

\newcommand\nat{Nature}

\Title{Multi-Wavelength Study of a Proto-BCG at z = 1.7}

\TitleCitation{Multi-Wavelength Study of a Proto-BCG at z = 1.7}

\Author{Quirino D'Amato $^{1,2,}$*\orcidA{}, Isabella Prandoni $^{1}$, Marisa Brienza $^{1,2}$, Roberto Gilli $^{3}$, Cristian Vignali $^{1,2}$, \mbox{Rosita Paladino $^{1,4}$}, Francesca Loi $^{5}$, Marcella Massardi $^{1,4}$, Marco Mignoli $^{3}$, Stefano Marchesi $^{3}$, Alessandro Peca $^{6}$ \mbox{and Preshanth Jagannathan $^{7}$}}

\AuthorNames{Quirino D'Amato, Isabella Prandoni, Marisa Brienza, Roberto Gilli, Cristian Vignali, Rosita Paladino, Francesca Loi, Marcella Massardi, Marco Mignoli, Stefano Marchesi, Alessandro Peca and Preshanth Jagannathan}

\AuthorCitation{D'Amato, Q.; 
Prandoni, I; Brienza, M.; Gilli, R.; Vignali, C.; Paladino, R.; Loi, F.; Massardi M.; Mignoli, M.; Marchesi, S.; et al.}

\address{%
$^{1}$ \quad INAF/IRA, Istituto di Radioastronomia, Via Piero Gobetti 101, 40129 Bologna, Italy; \mbox{prandoni@ira.inaf.it (I.P.);} 
 m.brienza@ira.inaf.it (M.B.); cristian.vignali@unibo.it (C.V.); \mbox{rosita.paladino@inaf.it (R.P.)}; massardi@ira.inaf.it (M.M.)\\
$^{2}$ \quad Dipartimento di Fisica e Astronomia dell’Università degli Studi di Bologna, via P. Gobetti 93/2, \mbox{40129 Bologna, Italy}\\
$^{3}$ \quad INAF/OAS, Osservatorio di Astrofisica e Scienza dello Spazio di Bologna, via P. Gobetti 93/3, \mbox{40129 Bologna, Italy}; roberto.gilli@inaf.it (R.G.); marco.mignoli@inaf.it (M.M.); \mbox{stefano.marchesi@inaf.it (S.M.)}\\
$^{4}$ \quad Italian ALMA Regional Center (ARC), Via Piero Gobetti 101, 40129 Bologna, Italy\\
$^{5}$ \quad INAF-Osservatorio Astronomico di Cagliari, Via della Scienza, 5, 09047 Cuccuru Angius, Italy; francesca.loi@inaf.it\\
$^{6}$ \quad Department of physics, University of Miami, Coral Gables, FL 33124, USA; alessandro.peca@miami.edu\\
$^{7}$ \quad National Institute of Radio Astronomy (NRAO) 801 Leroy Place,
 Socorro, NM 87801, USA; pjaganna@nrao.edu}

\corres{Correspondence: quirino.damato2@unibo.it}

\abstract{\textls[-5]{In this work we performed a spectral energy distribution (SED) analysis in the optical/infrared band of the host galaxy of a proto-brightest cluster galaxy (BCG, NVSS \mbox{J103023+052426})} \textls[15]{in a proto-cluster at z = 1.7. We found that it features a vigorous star formation rate (SFR) of \mbox{${\sim}$570 $\mathrm{M_{\odot}}$/yr} and a stellar mass of $M_{\ast} \sim 3.7 \times 10^{11}$ $\mathrm{M_{\odot}}$; the high corresponding specific} \mbox{SFR = $1.5 \pm 0.5$ $\mathrm{Gyr^{-1}}$} classifies this object as a starburst galaxy that will deplete its molecular gas reservoir in $\sim$$3.5 \times 10^8$ yr. Thus, this system represents a rare example of a proto-BCG caught during the short phase of its major stellar mass assembly. Moreover, we investigated the nature of the host galaxy emission at 3.3 mm. We found that it originates from the cold dust in the interstellar medium, even though a minor non-thermal AGN contribution cannot be completely ruled out. Finally, we studied the polarized emission of the lobes at 1.4 GHz. We unveiled a patchy structure where the polarization fraction increases in the regions in which the total intensity shows a bending morphology; in addition, the magnetic field orientation follows the direction of the bendings. We interpret these features as possible indications of an interaction with the intracluster medium. This strengthens the hypothesis of positive AGN feedback, as inferred in previous studies of this object on the basis of X-ray/mm/radio analysis. In this scenario, the proto-BCG heats the surrounding medium and possibly enhances the SFR in nearby galaxies.} 


\begin{document}

\section{Introduction}
\label{sec:intro}
Galaxy clusters are the gravitationally bound structures in the Universe, key environments in which to trace the formation of the most massive dark matter halos, galaxies and super massive black holes (SMBHs). Numerical simulations invoke feedback from active galactic nuclei (AGN) to reproduce the observed properties of massive galaxies in the local Universe \citep{antonuccio_2008,gaibler_2012} and clear signatures of jet-induced AGN feedback are indeed observed in local galaxy clusters (\cite{salome_2016}, and references therein). AGN feedback may be even more important at earlier stages of galaxy evolution, particularly during the so-called ``cosmic noon'' (z${\sim}$2), when both AGN and star formation activities peak \citep{gruppioni_2011, delvecchio_2014, rodighiero_2015}. Among~distant AGN, high-z radio galaxies (HzRGs) are found to be excellent protocluster signposts up to z${\sim}$4 \citep{venemans_2007,miley_2008,chiaberge_2010}. In~these systems the AGN radio jets inject enormous amounts of energy in the intracluster medium (ICM) and, according to simulations~\cite{fragile_2017}, can even enhance star-formation in the galaxies nearby. Nevertheless the mechanisms behind the interaction between the several components inproto-clusters, such as galaxies and the hot ICM, are still poorly investigated, especially prior to virialization \citep{overzier_2016}.

\subsection*{The J1030 Protocluster: Summary of Previous~Results}

\begin{figure}[H]	
\widefigure
\includegraphics[width=.92\textwidth,keepaspectratio]{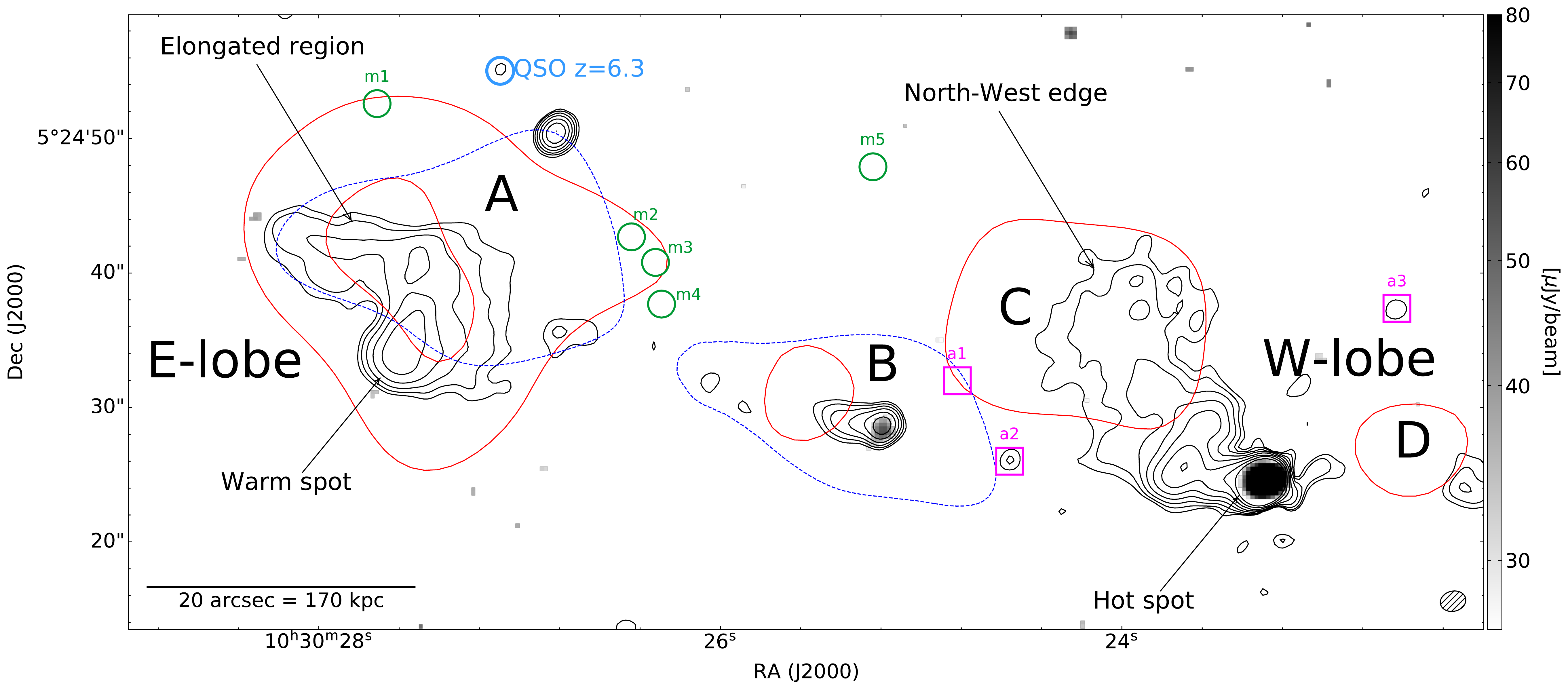}
\caption{Image: ALMA continuum emission at 92.04 GHz, blanked at the 3$\sigma$ level. The~black contours indicate the total intensity emission of the FRII HzRG at 1.4 GHz, starting at the ${\sim}3 \sigma$ level emission and increasing with a $\sqrt{3}$ geometric progression. Blue (dashed) and red (solid) contours are the 2--3$\sigma$ soft and 2$\sigma$ hard X-ray diffuse emission, respectively. The~four major components of such emission are labeled in black (\textbf{A}--\textbf{D}). The~positions of the sources discovered by MUSE ($m1-m5$) are indicated by the green circles (a sixth source is located outside the image, ${\sim}$45 arcsec north-east from the FRII core). The~positions of the protocluster members discovered by ALMA ($a1-a3$) are marked by the magenta boxes. The~Eastern and Western lobes are labeled as E-lobe and W-lobe, respectively. The~main morphological features of the lobes are also reported, indicated by the black arrows. The~light blue circle indicates the z = 6.3 QSO (i.e., the~center of the J1030 field). The~solid black line at bottom-left corner shows the angular and physical scale, while the dashed ellipse at the bottom-right corner is the restoring beam of the ALMA image.
\label{fig:alma_cont}}
\end{figure}

A remarkable structure with which to investigate the role of AGN feedback in regulating  ICM heating and star formation is the z = 1.7 proto-cluster located in the so-called ``J1030 field'', centered around the $z = 6.3$ Sloan Digital Sky Survey (SDSS) Quasar (QSO) J1030+0524 \citep{fan_2001}. This structure comprises at least 10 star-forming galaxies (SFGs) distributed around a powerful Fanaroff--Riley type II (FRII; \cite{fanaroff_1974}) HzRG at RA=$10^h 30^m 25^s.2$, Dec=$+5^{\circ} 24' 28''.7$, part of the  NRAO VLA Sky Survey (NVSS, ID: J103023+052426; \cite{condon_1998}) and hosting a heavily obscured AGN (X-ray derived column density $N_{H,X}{\sim}10^{24}~\mathrm{cm^{-2}}$; \cite{gilli_2019}). Seven SFGs have been spectroscopically confirmed through  Very Large Telescope Multi Unit Spectroscopic Explorer (VLT/MUSE) and Large Binocular Telescope Utility Camera in the Infrared (LBT/LUCI) observations \citep{gilli_2019}. The~position of five VLT/MUSE sources is indicated by green circles in Figure~\ref{fig:alma_cont}; a sixth source is located ${\sim}$40 arcsec north from the FRII core, while a source detected by LBT/LUCI is located ${\sim}$1.5 arcmin south-east, both outside the image. Based on Atacama Large Millimeter Array (ALMA) observations of the CO(2--1) transition line, ref.~\cite{damato_2020b} reported the discovery of three new gas-rich members (magenta circles in Figure~\ref{fig:alma_cont}), in~addition to a large molecular gas reservoir ($M_{H_2} \sim 2 \times 10^{11}$ $\mathrm{M_{\odot}}$) distributed in a ${\sim}$20--30 kpc structure around the FRII host galaxy and rotating perpendicularly to the radio jets (at least in projection). Furthermore, ref.~\cite{damato_2020b} demonstrated that the overdensity will collapse in a cluster of at least $10^{14}$ $\mathrm{M_{\odot}}$ at z = 0, and~that the FRII galaxy will likely evolve into the future brightest cluster galaxy (BCG). Ref. \citep{gilli_2019} also reported the discovery of an X-ray diffuse emission detected by \textit{Chandra} in both the soft and hard bands, composed of four major components (Figure~\ref{fig:alma_cont} A--D; the soft and hard ${\sim}$2$\sigma$ emissions are indicated by the red and blue contours, respectively). The~component B is co-spatial with the FRII core and jet-base, and~is well-fitted by a flat-spectrum non-thermal model (photon index $\Gamma \sim 0.1$), while the C and D components are ascribed to thermal emission ($T_{gas} \sim 1$ keV). As~for the A component, the~emission likely arises from a combination of thermal (gas heating) and non-thermal (inverse Compton with the cosmic microwave background) processes (see~\cite{gilli_2019} for further details). Remarkably, four SFGs ($m1-m4$ in Figure~\ref{fig:alma_cont}) lie in an arc-like shape around the edge of the A component. Moreover, these galaxies show a specific star formation rate (sSFR) higher than that of the other protocluster members. Ref.~\cite{gilli_2019} proposed that the X-ray diffuse emission originates from an expanding bubble of gas, shock-heated by the energy inflated by the FRII jet into the ICM. This could be the first evidence of AGN feedback enhancing the SFR in nearby galaxies at distances of hundreds of kpc. Since~2018, the~structure has been targeted by deep Jansky Very Large Array (JVLA) observations at \mbox{1.4 GHz} (D’Amato~et~al., subm., ref. \cite{damato_subm}) that have revealed extended radio emission around the lobes of the FRII galaxy (green contours in Figure~\ref{fig:alma_cont}). 
The Western lobe (W-lobe) features a clear hot spot in the total intensity map, in~addition to an extended emission the North-East direction, that overlaps with the X-ray component C. The~Eastern lobe (E-lobe) lacks of a classical hot spot (i.e., brightness contrast with the rest of the radio source $\geq4$, following the definition of \citep{deruiter_1990}) and shows a complex structure, with~the peak emission concentrated at a warm spot and an elongated structure along the Northern boundary. Finally, the~protocluster has been recently observed at 150 MHz with the Low Frequency Array (LOFAR), showing the presence of a more extended emission around the FRII lobes owing to low-energy electrons (Brienza~et~al., subm., ref. \cite{brienza_subm}). Coupled with the JVLA observations, the~LOFAR data allowed us to build-up a spectral index map that unveils a spectral index flattening towards the external regions of the lobes, beyond~the eastern and western spots (Brienza~et~al., subm., ref. \cite{brienza_subm}).
In this work we present a multiband analysis of the FRII host galaxy to characterize its stellar mass, SFR and AGN bolometric luminosity. Furthermore, we analyze the polarized emission at 1.4 GHz as observed with the JVLA to look for possible signatures of interactions between the radio galaxy and the ICM. In~Section~\ref{sec:observations} we describe the data used in this work, the~data reduction and the imaging process. In \mbox{Section~\ref{sec:results}} we analyze the data and discuss the implications of our findings for the FRII host galaxy and the FRII positive feedback scenario. Finally, in~Section~\ref{sec:conclusions} we summarize our results and we draw our~conclusions. 

\textls[20]{Throughout the work we adopt a concordance $\Lambda$CDM cosmology with} $\mathrm{H_0}$ = \textls[-5]{\mbox{70 $\mathrm{km~s^{-1}}$ $\mathrm{Mpc^{-1}}$}, $\Omega_{\mathrm{M}} = 0.3$, and~$\Omega_{\Lambda} = 0.7$, in~agreement with the \textit{Planck 2015} \mbox{results \citep{PLANCK_2016}}}.
The angular scale and luminosity distance at z = 1.7 are 8.46 kpc/arcsec and 12.7 Gpc, respectively.
\end{paracol}
\nointerlineskip
\begin{paracol}{2}

\switchcolumn
 \vspace{-12pt}
\section{Observations and Data~Reduction}
\label{sec:observations}
\unskip
\subsection{Broadband Optical/IR~Photometry}

The J1030 field features exceptional multiband coverage. About
0.15 $\mathrm{deg^2}$ around the z = 6 QSO have been imaged in both optical and near-IR bands. It is part of the Multi-wavelength Yale--Chile survey (MUSYC; ref.~\cite{gawiser_2006, blanc_2008}) and Canada--France Hawaii Telescope Wide-field InfraRed Camera (CFHT/WIRCam; ref.~\cite{balmaverde_2017}) observations, which provide photometry in the \textit{UBVRIzJHK} and \textit{YJ} bands, respectively. It is also covered by the \textit{Spitzer} Infrared  array Camera (IRAC) and Multiband Imaging Photometer for Spitzer (MIPS) mosaics~\cite{annunziatella_2018}, and~\textit{Hershel} Photodetecting Array Camera and Spectrometer (PACS) and Spectral and Photometric Imaging Receiver (SPIRE) observations~\cite{leipski_2014}. The~innermost region is covered by deep Hubble Space Telescope (HST) Advanced Camera for Surveys (ACS) and Wide Field Camera 3 (WFC3) observations (\cite{stiavelli_2005,damato_2020b}). 
In this work we exploit available photometry to perform a spectral energy distribution (SED) analysis of the FRII host galaxy (Section \ref{sec:SED_fitting}). Since the FRII optical/IR emission is located in proximity to a bright star, the~archive photometry in several bands (especially at short wavelengths) suffers from contamination. Thus, in~previous works we re-analyzed the data for HST and MUSYC. As~for the WIRCam and IRAC channels, we followed the same photometry measurements as described in \citep{balmaverde_2017}.
The measured fluxes and their uncertainties are reported in Table~\ref{tab:SED_points}. 

    \begin{specialtable}[H]
    \tablesize{\small}
\caption{Photometric data-points used to perform the SED fitting. Column 1: observing wavelength. Column 2: observed flux density. Column 3: observing instrument. Column 4: archive references, where we specify whether the data have been re-analyzed. \label{tab:SED_points}}
\setlength{\cellWidtha}{\columnwidth/4-2\tabcolsep-0.5in}
\setlength{\cellWidthb}{\columnwidth/4-2\tabcolsep-0in}
\setlength{\cellWidthc}{\columnwidth/4-2\tabcolsep-0in}
\setlength{\cellWidthd}{\columnwidth/4-2\tabcolsep+.5in}
\scalebox{1}[1]{\begin{tabularx}{\columnwidth}{>{\PreserveBackslash\centering}m{\cellWidtha}>{\PreserveBackslash\centering}m{\cellWidthb}>{\PreserveBackslash\centering}m{\cellWidthc}
>{\PreserveBackslash\centering}m{\cellWidthd}}
\toprule
\boldmath{$\lambda$} \textbf{[}\boldmath{$\upmu$}\textbf{m]}	& \boldmath{$S_\nu$} \textbf{[}\boldmath{$\upmu$}\textbf{Jy]}	& \textbf{Instrument} & \textbf{Reference}\\
\midrule
    0.9	 & $1.6 \pm 0.2$      & HST/ACS        & \cite{stiavelli_2005}, reanalyzed in~\cite{damato_2020b} \\
    1.2	 & $3.6 \pm 0.7$      &	CFHT/WIRCam    & \cite{balmaverde_2017}  (reanalyzed)            \\
        2.1	 & $16  \pm 2$	      & CTIO/ISPI     & \cite{quadri_2007}, reanalyzed in~\cite{gilli_2019}   \\
    3.5	 & $36  \pm 7$	      & Spitzer/IRAC   &  \cite{annunziatella_2018} (reanalyzed, see also~\cite{balmaverde_2017})        \\
    4.5	 & $58  \pm 12$	      & Spitzer/IRAC   &  ''        \\
    5.7	 & $100 \pm 20$	      & Spitzer/IRAC   &  ''        \\
    7.8	 & $76  \pm 15$	      & Spitzer/IRAC   &  ''        \\
   23.5	 & $620  \pm 62$      &	Spitzer/MIPS   & IRSA archive \\
  105.4	 & $7655  \pm 3274$   &	Herschel/PACS  & \cite{leipski_2014} \\
  169.5	 & $<$30,000	      & Herschel/PACS  & '' \\
  246.7	 & 33,400 $\pm$ 9800   &	Herschel/SPIRE & '' \\
  348.7	 & 43,600 $\pm$ 12,600  & Herschel/SPIRE & '' \\
  495.3	 & 36,100 $\pm$ 15,000  & Herschel/SPIRE & '' \\
 1120.5	 & $2500 \pm 500$     &	AzTEC & \cite{zeballos_2018} \\

\bottomrule
\end{tabularx}}
\end{specialtable}

\subsection{ALMA~Photometry}
We derived 3.3-mm photometry from proprietary ALMA observations, composed by a mosaic of three pointings performed in Band 3 (84--116 GHz) during Cycle 6 (project ID: {2018.}1.{01601}.S, PI: R. Gilli). Details about the scheduling blocks and calibrators can be found in~\cite{damato_2020b}.
The $2 \times 4$ GHz spectral windows have been set to cover the frequency ranges 84.1--87.7 GHz and 96.1--99.9 GHz, respectively. Each spectral window consists of 1920 channels of 976.5 kHz width. The~data calibration has been performed using the calibration pipeline of the Common Astronomy Software Applications (CASA) package (version 5.4.0-70; \citep{mcmullin_2007}).

The observations mainly aimed at detecting the CO(2--1) transition line in the protocluster members. Refs.~\cite{gilli_2019} and~\cite{damato_2020} showed that the protocluster members can be found in a wide velocity range (${\sim}$1500 km/s) around the mean redshift structure (z = 1.694). Thus, in~order to avoid any possible contamination from line emission in the field, we performed the continuum imaging in the available bandwidth of the observations, excluding a range of $\pm$1500 km/s from the mean redshift of the protocluster. The~imaging was performed with the {\sc{tclean}} CASA task in multi-frequency synthesis (\textit{mfs}) mode, resulting in an observed-frame frequency of the image equal to 92.04 GHz. We used a Briggs weighting scheme with robustness parameter equal to 0.5, corresponding to a restoring beam with a major (minor) axis of 1.92 (1.50) arcsec. The~average root mean square (\textit{rms}) within the primary beam half maximum (PBHM) region is ${\sim}$9 $\upmu$Jy. The~continuum emission at the 3$\sigma$ level is shown in Figure~\ref{fig:alma_cont}.


\subsection{JVLA Observations of Polarized~Emission}

We performed 11 observations of the J1030 field with the JVLA in the L-band (1--2 GHz, project ID: VLA/18A-440, PI: I. Prandoni). Details about the observation and total intensity calibration can be found in D'Amato~et~al. (subm., ref. \cite{damato_subm}). The~dataset was found to be strongly affected by radio--frequency interference (RFI), as~expected for the JVLA L-band 
\mbox{(\url{https://science.nrao.edu/facilities/vla/docs/manuals/obsguide/rfi}}, accessed on 6 December 2021
); this required a careful manual calibration, as~well as heavy data-flagging and four of the \mbox{11 observing blocks} were entirely discarded. Here, we focus on polarization~calibration.

\subsubsection{Instrumental~Calibration}
The first step is to determine the instrumental delay and leakage between the two polarizations for each dataset, before~calibrating the position angle using a source with a known polarization position angle. Standard polarization calibration normally requires the observation of a polarization calibrator. While the leakage terms can be calculated using a non-polarized source at the observed frequency, a~source with known polarized emission is required to correct for the instrumental cross-hand delay and calibrate the position angle. Unfortunately, we did not observe a polarized calibrator, since a polarization analysis was not in the original intent of the observations. However, one of the observed calibrators (3C147) is unpolarized in the L-band, so it is possible to use it as leakage calibrator. In~addition, if~a significant leakage emission is detected in the linear polarization image of the unpolarized calibrator, then the leakage signal itself can be used to correct for the delays. We performed a linear polarization imaging of 3C147, measuring a polarized emission equal to ${\sim}$1\% of the total intensity, significantly higher than the expected upper limit reported for 3C147 ($<$0.05\%; 
\url{https://science.nrao.edu/facilities/vla/docs/manuals/obsguide/modes/pol}, accessed on 6 December 2021
). Thus, we initially calculated the cross-hand delays using the {\sc{gaincal}} task of CASA, and,~subsequently, we calculated the leakage terms, applying, on the fly, the cross-hand delays table. Finally we applied the leakage and delay tables to the target (J1030) and calibrator (3C147) fields. We re-imaged the total intensity and linear polarization of 3C147 after the corrections, finding that the total intensity was conserved, while the polarized emission was strongly reduced (fractional polarization ${\sim}$0.01\%, consistent with the tabulated value of $<$0.05\%).

\subsubsection{Polarization~Calibration}
Once the datasets were corrected for the instrumental effects, we needed to calibrate the polarized emission and polarization angle. Although~this requires the observation of a polarized calibrator, we, fortunately, found a bright polarized source (signal-to-noise ratio \mbox{S/N $>$ 1000}, flux density $S_\nu \sim 200$ mJy) in our target field that is part of the NVSS. The~mean polarized fraction ($\overline{F}\sim 4.73\pm0.3\%$) and mean polarization angle (\mbox{$\overline{\tau}=6.8\pm0.9$ deg}) derived from the averaging of the two NVSS intermediate frequencies (IFs, 42 MHz bandwidth centered at 1365 MHz -- IF1 -- and 1435 MHz -- IF2) of this source have been reported by \cite[][]{condon_1998}, while the rotation measure ($RM$ = 11.7 $\pm$ 7.5 $\mathrm{deg/m^2}$) has been reported by~\cite{taylor_2009}. Exploiting the value of $\overline{\tau}$ and the relation between the $RM$ and the change in the
position angle between the two IFs ($\Delta \tau$, Equation~(1) in~\cite{taylor_2009}), we derived the two position angles in the IF1 ($\tau_1 = 8.4$ deg) and IF2 ($\tau_2= 5.2$ deg) of J102921+051938. We note that we assumed no wrapping of the magnetic field between the IF1 and IF2, as~appropriate for sources with an $RM <$ 400 $\mathrm{deg/m^2}$ \citep{taylor_2009}.
As for the polarization fraction in the two IFs, we assumed the value of $\overline{F}$. Considering the IFs are contiguous and rather narrow, we expect that assuming an average value provides estimates of the polarized emission in the two IFs accurate enough for our purposes. Once that we had derived the polarization parameters of J102921+051938 in the two IFs, we shifted the phase center of the target field on this calibrator, obtaining the new calibrator field. Thus, we split our JVLA dataset in two subsets, identical to the NVSS IFs. For~each subset, we set the polarized emission model for the calibrator field accordingly to the fractional polarization and polarization angles derived for J102921+051938 (using the {\sc{setjy}} task). Finally, we calculated the gains from the calibrator field and applied them to the target field. In~order to check whether the polarized fraction and polarization angle had been correctly applied, we performed a quick imaging of J102921+051938 in the target field, and~measured the polarized emission (from the linear polarization image, i.e.,~$\sqrt{U^2+Q^2}$, where $U$ and $Q$ are the Stokes parameters) and polarization angle, finding a good agreement ($<$1$\sigma$) with the model~values.

We performed the target (i.e., the~FRII HzRG) imaging of the Stokes $I$, $U$ and $Q$, separately for the two IFs subsets, with~a robustness parameter of 0.5; we applied a mild tapering of $<$50 k$\lambda$ (corresponding to a ${\sim}$2.5 arcsec restoring beam), in~order to retrieve the extended emission with sufficient resolution to distinguish the different components in the FRII galaxy lobes. We tested several bandwidths (5, 10, 50 and 100 MHz) centered at 1.4~GHz on the uncalibrated dataset, in~order to check at which bandwidth the depolarization became significant, vanishing the gain in sensitivity obtained from widening the frequency range. We found that the bandwidth depolarization started to decrease the S/N of the polarized flux between 50 MHz and 100 MHz; thus, in~the two 42 MHz IFs bandwidth, depolarization can be considered negligible. The~debiased linear polarization image is calculated as $\sqrt{U^2+Q^2-2.3\sigma_{UQ}^2}$, where $\sigma_{UQ}$ is the average \textit{rms} of the $U$ and $Q$ images, following Equation~(5) of~\cite{george_2012}. This formula is suited for S/N $\geq$ 4 polarized emission. In~our map, the only regions above S/N = 4 are the hot and warm spots, while the rest of the emission has S/N $\geq$ 3. Despite this implying that the absolute value of fractional polarization is less reliable in the low-S/N regions, we stress that the debias applies a general offset to all pixels. Our analysis aims at showing an increasing of the fractional polarization in the extended regions with respect to the hot and warm spots, so this offset does not affect our results. The target polarization maps in the IF1 and IF2 have a sensitivity of 36 $\upmu$Jy and 32 $\upmu$Jy, respectively. In~each IF, we computed the fractional polarization map as the ratio between the debiased linear polarization and the total intensity maps. In~Figure~\ref{fig:polar} we report the fractional polarization map of the FRII (blanked at the 3$\sigma$ level of the linear polarization) in the IF1 (top) and IF2 (bottom). The~magnetic field orientation is calculated as $0.5\arctan(U/Q)+\pi/2$, in~steps of three pixels, and~is indicated by the cyan~dashes.

\end{paracol}
\nointerlineskip
\begin{figure}[H]	
\widefigure
\includegraphics[width=.95\textwidth,keepaspectratio]{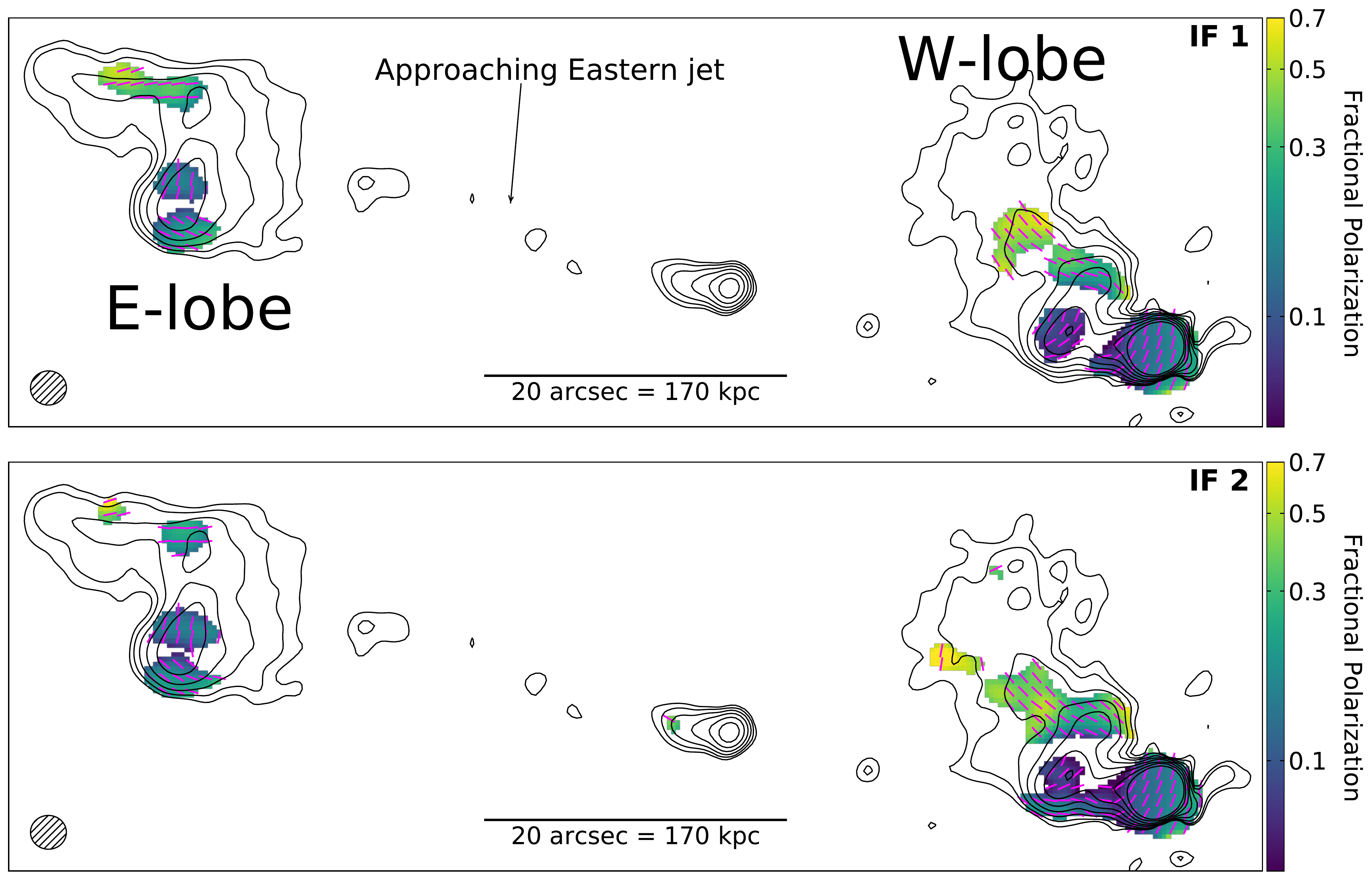}
\caption{Fractional polarization images of the FRII HzRG in the two 42-MHz-wide IFs: IF1 (1365~MHz, \textbf{top}) and IF2 (1435~MHz, \textbf{bottom}), blanked at the 3$\sigma$ level. The~black contours indicate the total intensity emission of the high-resolution, full-band image, starting at the ${\sim}3 \sigma$ level emission and increasing with a $\sqrt{3}$ geometric progression. The~magenta dashes indicate the orientation of the magnetic field. The~dashed ellipse in the bottom-left corner represents the restoring beam. The~solid black line in the bottom is the angular and physical scale. For~clarity, in~the IF1 image we indicated the Eastern and Western lobes as E-lobe and W-lobe, respectively, and~specify that the Eastern jet is the approaching one.
\label{fig:polar}}
\end{figure}  
\begin{paracol}{2}

\switchcolumn

\section{Results and~Discussion}
\label{sec:results}
\unskip

\subsection{SED Fitting: A Proto-BCG Caught at the Peak of Its Stellar Mass~Building}
\label{sec:SED_fitting}

We modeled the SED of the FRII host galaxy emission using the fitting code originally presented by~\cite{fritz_2006} and improved by~\cite{feltre_2012}. The~code fits the data, simultaneously accounting for three different components: stellar emission, modeled by means of simple stellar populations (SSPs); reprocessed emission from the dusty torus surrounding the AGN; and~emission by the cold dust of the host galaxy that is heated by starburst activity. The fitting code and the libraries used to model the SED are the same ones adopted by~\cite{circosta_2019} and are described in details in Section~5.2 of their work. In~addition, we performed a second SED fitting with the same models except for the cold dust contribution, which has been fitted with a simple gray-body model with $\beta=2$ (as generally suited for high-z, gas-reach-obscured AGN, refs.~\cite{conley_2011,fu_2012,riechers_2013,gilli_2014} and local starburst-gas-rich starburst galaxies, ref. ~\cite{rangwala_2011}). In~both fits, we excluded the ALMA data point, due to the uncertain nature of the ALMA continuum emission (i.e., either due to thermal or synchrotron emission, see Section \ref{sec:alma_results}). The~reduced $\chi^2$ is equal to 1.2 and 1.0 for the fitting using the cold dust template and the simple gray-body model, respectively.
In Figure~\ref{fig:SED_fitting} we report the SED of the FRII host galaxy and the fitting model (black solid line), as~well as the three emission components of the fit. The~left panel shows the SED fitting using the cold dust template, while the right panel shows the SED fitting where the dust emission has been modeled using the~gray-body.

\begin{figure}[H]
\includegraphics[width=0.73\textwidth,keepaspectratio]{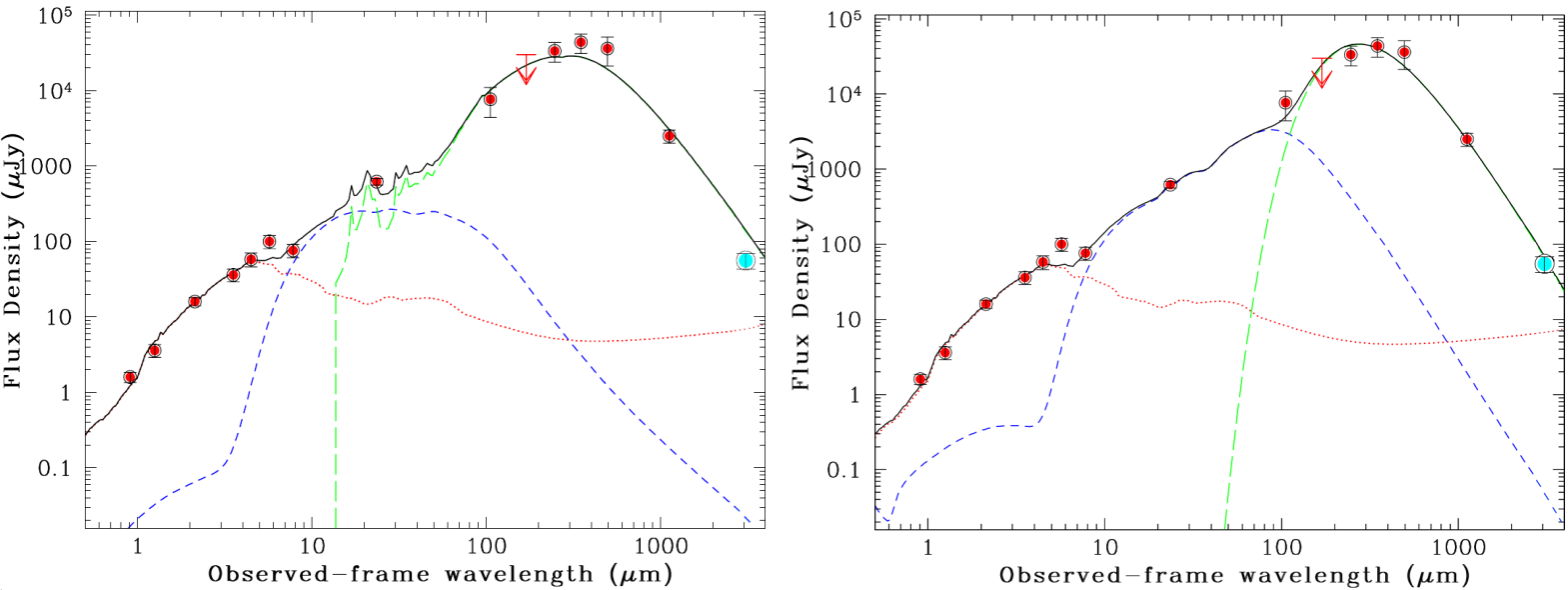}
\caption{SED fitting of the FRII host galaxy. Photometric points are marked by the filled red circles, while the total fitting model is the solid black line. In~both panels the ALMA point is not included during the fitting and is marked by the filled cyan circle. (\textbf{Left panel}): The model includes three components: stellar emission (indicated by the red dotted line), the~reprocessed emission of the dusty torus surrounding the AGN (blue dashed line), and~emission by the cold dust of the host galaxy, heated by starburst activity (green dashed line). (\textbf{Right panel}): same as the left panel, where only the cold dust template has been substituted by a simple gray-body model with $\beta=2$.
\label{fig:SED_fitting}}
\end{figure}

The fiducial errors are 20\% for the bolometric and far-infrared (FIR) luminosity, and~30\% for the stellar mass; these uncertainties are based on the error analysis of~\cite{circosta_2019}, who take into account all acceptable solutions within 1$\sigma$ confidence level, that ism within a given range of $\Delta \chi^2$ that depends on the free parameters of the SED-fitting procedure, and~by comparison between different fitting codes (see Section~5.2 of~\cite{circosta_2019} for caveats and details).
The AGN fraction in the 8--1000 $\upmu$m range of the gray-body model fitting is significantly higher (24\%) that that derived from the dust-template fitting (3.1\%), even if the total FIR luminosity is comparable ($L_{8-1000~\upmu\mathrm{m}} \sim 3.2 \times 10^{12}$ $\mathrm{L_{\odot}}$ for the dust-template and $L_{8-1000~\upmu\mathrm{m}} \sim 3.7 \times 10^{12}$ $\mathrm{L_{\odot}}$ for the gray-body, respectively), due to the higher contribution of the dust-template at shorter wavelengths with respect to the gray-body.
This high fraction derived from the gray-body model fitting is more consistent with what is expected for Type II heavily obscured AGN, where the torus absorbs and re-emits most of the light originated in the accretion disk \citep{padovani_2017}; indeed, our source is Compton thick and~\cite{gilli_2019} derived an edge-on orientation ($80^\circ$) of the FRII radio-galaxy with respect to the line of sight. 
In addition, once the AGN fraction is subtracted, the~FIR luminosity difference is ${\sim}$10\%, that is within the 1$\sigma$ uncertainty; thus, the~SFRs derived exploiting the $L_{\mathrm{FIR}}$ --
SFR linear calibration presented by~\cite{kennicutt_1998} and subtracting the AGN fraction are in agreement within 1$\sigma$ for the two fittings (\mbox{570--630 $\mathrm{M_{\odot}}$/yr}). 
As for the AGN bolometric luminosity derived from the SED fitting, we obtain \mbox{$L_{\mathrm{bol,~AGN}} \sim 1.5 \times 10^{46}$ erg/s} and $L_{\mathrm{bol,~AGN}} \sim 1.1 \times 10^{46}$ erg/s for the dust-template and gray-body, respectively. Adopting, as~bolometric corrections, the~values derived by~\cite{duras_2020} (see also~\cite{lusso_2012}) for a large samples of AGN spanning seven dex in luminosity, we predict an intrinsic 2--10 keV luminosity of $L_{\mathrm{2-10 keV}} \sim 3.8 \times 10^{44}$ erg/s and $L_{\mathrm{2-10 keV}} \sim 3.2 \times 10^{44}$ erg/s, respectively, which are a factor of ${\sim}$3--2.5 higher that the one reported by~\cite{gilli_2019} by fitting the X-ray spectrum. Given all the uncertainties (just to mention two 0.27 dex dispersion in~\cite{duras_2020} relation and the correction for the column density applied in~\cite{gilli_2019}), we may conclude that the AGN luminosity derived from the SED-fitting is broadly consistent with that derived using X-rays, especially in the case of the gray-model body.
From the two SED fits we derived consistent (1$\sigma$) AGN bolometric luminosity, FIR luminosity, stellar mass and SFR. Given the aforementioned considerations about the AGN luminosity and the slightly better $\chi^2=1.0$, we report here the values derived from the SED fitting with the gray-body model, stressing the assumption that the values of stellar mass and SFR derived from the dust-template fitting would not affect the following conclusions about the starburst nature of the galaxy. We derive: \mbox{$L_{\mathrm{bol,~AGN}} \sim 1.1 \times 10^{46}$ erg/s}, $L_{8-1000~\upmu\mathrm{m}} \sim 3.7 \times 10^{12}$ $\mathrm{L_{\odot}}$ (24\% of which is ascribed to the AGN), stellar mass $M_{\ast} \sim 3.7 \times 10^{11}$ $\mathrm{M_{\odot}}$ and \mbox{SFR = 570 $\mathrm{M_{\odot}}$/yr.}
The high \mbox{sSFR = $1.5 \pm 0.5$ $\mathrm{Gyr^{-1}}$} derived for the FRII host galaxy classifies this object as a starburst galaxy when compared with other SFGs at the same \mbox{redshift \citep{schreiber_2015}}, implying that the proto-BCG is currently being observed during the brief phase (few $10^8$ yr; ref.~\cite{lapi_2018}) of its major stellar mass building. Indeed, we note that the derived SFR implies that the molecular gas reservoir measured by \citep{damato_2020b} will be depleted in ${\sim}3.5 \times 10^8$ yr. Moreover, assuming that the BH is accreting with an efficiency $\eta=0.3$ (as assumed by~\cite{gilli_2019} and as generally thought for BHs powering jetted AGN; refs.~\cite{blandford_1977,ghisellini_2014}), from~the bolometric luminosity, we derive an accretion rate for the SMBH of \mbox{$\dot{M} = L_{\mathrm{bol,~AGN}}/(\eta c^2) \approx 0.6 $ $\mathrm{M_{\odot}}$/yr.}

\subsection{Continuum Emission at 3.3~mm}
\label{sec:alma_results}
In the ALMA continuum map at 3.3 mm, we detected the FRII core and the Western hot spot at $>$3$\sigma$ level (Figure \ref{fig:alma_cont}). The~measured flux density of the core is $55 \pm 9$ $\upmu$Jy. This measurement is nicely in agreement with the flux density expected from the dust thermal emission, considering the best fit using the gray-body model (1$\sigma$), while it is incompatible at the 3$\sigma$ level with the expected flux density in the case of dust-template fitting, which is $\times$2.4 higher. Considering the best $\chi^2$ and the more plausible AGN fraction of the gray-body model (Section \ref{sec:SED_fitting}), this result strongly argues for thermal emission from the cold dust in the ISM as the origin of the observed flux density. In~addition, from~$L_{8-1000~\upmu\mathrm{m}}$ we can derive the expected flux density due to the star formation at \mbox{1.4 GHz} and compare it with our measurement in the JVLA map. Exploiting the stellar mass- and redshift-dependent $L_{8-1000~\upmu\mathrm{m}} - L_{1.4~\mathrm{GHz}}$ relation recently presented by~\cite{delvecchio_2021}, we derive a SFR-driven flux density  $S_{1.4~\mathrm{GHz, SFR}}{\sim}$90 $\upmu$Jy. This value is well below than that measured on the JVLA map ($S_{1.4~\mathrm{GHz}}=315 \pm 25$ $\upmu$Jy, ref.~\cite{damato_subm}), confirming that the flux density at 1.4 GHz is dominated by the AGN. If~we subtract $S_{1.4~\mathrm{GHz, SFR}}$ from the measured $S_{1.4~\mathrm{GHz}}$, and~we assume a wide synchrotron spectral index range $\alpha$ = 0.5--1.0 ($S_\nu \propto \nu^{-\alpha}$, valid from the jet base to the oldest electron population; refs.~\cite{giovannini_2001,ruffa_2019,padovani_2017}), we derive an expected synchrotron emission at 3 mm $S_{3~\mathrm{mm, sync}}$ = 4--30 $\upmu$Jy. Despite that the upper value (corresponding to $\alpha$ = 0.5) is barely compatible, within 3$\sigma$, with the flux density measured in our ALMA map, we argue that, even in this limited case with a flat slope, the expected flux is considerably lower than that measured. This, coupled with the SED-fitting results, strengthens the scenario in which the dust is primarily responsible for the observed flux density at 3 mm. However, we cannot completely rule out some minor contribution of synchrotron emission, due to the large uncertainties on the involved quantities (20\% on the $L_{8-1000~\upmu\mathrm{m}}$ and 30\% on the stellar mass), and~inherent scatter of the relation (${\sim}$0.2 dex, ref.~\cite{delvecchio_2021}) used to derive $S_{1.4~\mathrm{GHz, SFR}}$. In~addition, in~the Rayleigh--Jeans tail of the gray-body, the flux density is a function of the $\beta$ index, as $S_\nu \propto \nu^{2+\beta}$, where $\beta$ is assumed to be fixed at 2.0 during the fitting. In~summary, the~ALMA flux density is likely dominated by the dust emission, despite the fact that a minor non-thermal contribution may be present.

\subsection{Polarized Emission at 1.4 GHz: Signatures of AGN Feedback?}

The FRII HzRG shows a patchy structure in polarization (Figure \ref{fig:polar}). Both the western and eastern spots are detected, as~well as the terminal part of the western jet (clearly visible in IF2). Patchy polarized emission is also detected in the extended structures of the lobes. 
The fractional polarization in the W-lobe hot spot and E-lobe warm spot is in the 10--20\% range in both IFs. In~addition, the~magnetic field appears to be oriented along the jet axis in the western jet. Both the level of fractional polarization and the magnetic field orientation are consistent with that typically found in FRII radio galaxies \citep{parma_1993,muxlow_1991,pentericci_2000}.  
We note that the Eastern warm spot shows two regions of significant polarized emission; moreover, the~magnetic field seems to experience a 90-degree rotation going from the southern to the northern patch, showing a wrapping around the peak emission of the total intensity, as~observed in several cases around FRII hot spots \citep{saikia_1988, pentericci_2000}. We can then argue that the southern patch may correspond with the terminal region of the eastern jet, which then bends to the north, due to the impact with the ICM. The~polarized emission detected in the north--east 
elongated region of the E-lobe extends along the northern edge of the lobe and shows an enhanced polarization fraction ($\gtrsim$50\%), suggesting the presence of another compression front. In~this respect, Brienza~et~al. (subm., ref. \cite{brienza_subm}) found a flattening of the JVLA-LOFAR spectral index in the north--east elongated region, possibly ascribed to the shock re-acceleration of the electrons. This front may be related to the positive feedback scenario invoked by~\cite{gilli_2019} for the galaxies distributed along the northern edge of the X-ray component A (see Figure~\ref{fig:alma_cont}). 
In general, as~far as the extended regions of FRII lobes are concerned, we noticed an increasing of the polarization fraction and a co-spatiality of the polarized emission with regions where the total intensity emission shows a bending morphology. Also the orientation of the magnetic field seems to follow the observed bendings. Considering that the source is seen at an almost edge-on orientation with respect to the line of sight (${\sim}80^\circ$; \cite{gilli_2019,damato_subm}), 
these observational features seem, again, to suggest that these regions experience a compression and an enhancement of the magnetic field, possibly associated with the impact with components of the external medium. Indeed, the presence of gas bubbles overlapping with the radio galaxy lobes or located in their vicinity is testified by the detection of X-ray components (shown in Figure~\ref{fig:alma_cont}). 
Finally, we do not detect any significant polarized emission along the Northern edge of the W-lobe. In~this region~\cite{gilli_2019} reported the detection of an X-ray thermal diffuse emission (component C in Figure~\ref{fig:alma_cont}) with $T_{gas} \sim 1$ keV, which is the possible signature of shock-heated ICM. Moreover, Brienza~et~al. (subm. ref. \cite{brienza_subm}) showed a flattening of the spectral index moving towards this northern edge associated with the total intensity gradient along the same direction, strengthening the hypothesis of the presence of a shock front. However, we point out that the simplest explanation for the observed bending morphology in the lobes would be the peculiar motion of the galaxy through the ICM, if~evidence for the shock-heating mechanism found in previous and upcoming works \citep{gilli_2019,damato_2020,damato_subm} is neglected (a summary is given in Section~\ref{sec:intro}). In~addition the hot spots and knots may be partially optically thick, affecting the orientation of the observed magnetic field. Thus, our interpretation should be seen in the broad context of the these results, where the shock-induced ICM heating is able to simultaneously explain all the observed features at different wavelengths. Further observations at higher sensitivity of the polarization emission would ultimately confirm or reject this scenario, helping to shed light on the positive feedback~mechanism.


\section{Conclusions}
\label{sec:conclusions}
We analyzed the multi-wavelength properties of an FRII HzRG located at the center of a protocluster at z = 1.7 that is likely going to evolve into the future BCG. In~particular, we performed SED fitting of the FRII host galaxy, investigated the nature of the host galaxy emission at 3.3 mm, observed by ALMA, and~analyzed the polarized emission at 1.4 GHz, observed by the JVLA across the whole FRII galaxy morphology. Our conclusions are as follows:

\begin{enumerate}[leftmargin=2.3em,labelsep=0mm]
\item[-] Thanks to the SED fitting performed in the IR--mm band, we measured the physical parameters of the FRII host galaxy. From the best-fitting model we measured an AGN bolometric luminosity of $L_{\mathrm{bol,~AGN}} \sim 1.1 \times 10^{46}$ erg/s and a IR luminosity of $L_{8-1000~\upmu\mathrm{m}} \sim 3.7 \times 10^{12}$ $\mathrm{L_{\odot}}$ (24\% of which ascribed to the AGN), and~a stellar mass of $M_{\ast} \sim 3.7 \times 10^{11}$ $\mathrm{M_{\odot}}$. The~SFR corresponding to the AGN-subtracted IR luminosity is ${\sim}$570 $\mathrm{M_{\odot}}$/yr. The~SED-fitting bolometric AGN luminosity is consistent (\mbox{$\times$ 2.5}) with the X-ray-derived bolometric luminosity found by~\cite{gilli_2019}, given the large uncertainties involved. Considering the redshift of the source, the~high \mbox{sSFR = $1.5 \pm 0.5$ $\mathrm{Gyr^{-1}}$} unveils that the galaxy is experiencing a starburst phase in which it is assembling most of its final stellar mass. This represents a rare example of a proto-BCG caught in one of the most crucial phases of its~building-up.

\item[-] The flux density at 3.3 mm, measured with ALMA in the FRII core, is nicely in agreement (1$\sigma$) with the dust thermal emission expected from the best SED-fitting model, and~is significantly higher ($\times$ 1.8--13.7) than the expected SFR non-thermal emission as inferred from the 1.4 GHz emission. These findings strongly argue for a dominant thermal emission at the basis of the observed ALMA flux density. However, due to the large uncertainties in the derived quantities and exploited relations, a~minor non-thermal contribution cannot be completely ruled out.

\item[-] We detected polarized emission at 1.4 GHz in both FRII lobes. Both the eastern and western spots feature a fractional polarization of 10--20\% and a magnetic field perpendicular to the jet, as is~typically found for classical FRII. In general, we found an increased polarization fraction in the regions where the total intensity shows a bending morphology, and~a magnetic field orientation that seems to follow the direction of the bendings. Coupled with the X-ray diffuse emission detected in several spots around the FRII lobes (in part ascribed to shock-heated ICM, see~\cite{gilli_2019}), we interpret these features as possible signatures of compression produced by the external ICM. This strengthens the hypothesis of the positive AGN feedback scenario, wherein the AGN is responsible for the ICM heating, and~also, possibly, for the SFR enhancement in the SFGs located at the edge of the major component of the X-ray diffuse emission around the eastern lobe (see also~\cite{gilli_2019,damato_2020b}).\\
Further observations at higher sensitivities of the polarized emission will ultimately disclose the mechanisms at the origin of the observed features of the structure, and~probe how AGN feedback affects the thermodynamics of the ICM in this early structure and the physical properties of its member galaxies.
\end{enumerate}

\vspace{6pt} 



\authorcontributions{
Conceptualization, Q.D. and I.P.; Data curation, Q.D., I.P., M.B., C.V., F.L., M.M. (Marcella Massardi), M.M. (Marco Mignoli), S.M., P.J. and A.P.; Formal analysis, Q.D., I.P., M.B., C.V., R.P., M.M. (Marcella Massardi), M.M. (Marco Mignoli), S.M. and A.P.; Investigation, Q.D., I.P., M.B. and R.G.; Methodology, R.P., F.L. and M.M. (Marcella Massardi); Supervision, R.G., C.V., R.P., F.L. and M.M. (Marco Mignoli); Writing--original draft, Q.D.; Writing--review and editing, I.P., M.B., R.G., C.V., M.M. (Marcella Massardi), S.M., P.J. and A.P. All authors have read and agreed to the published version of the manuscript.
}

\funding{We acknowledge support from the agreement ASI-INAF n. 2017-14-H.O. I.P. acknowledges support from INAF under the SKA/CTA PRIN ``FORECaST'' and the PRIN MAIN STREAM ``SauROS'' projects. M.B. acknowledges financial support from the ERC-Stg 674 DRANOEL, no 714245 and the ERC Starting Grant ``MAGCOW", no. 714196. FL acknowledges financial support from the Italian Minister for Research and Education (MIUR), project FARE, project code R16PR59747, project name FORNAX-B. FL acknowledges financial support from the Italian Ministry of University and Research $-$ Project Proposal CIR01$\_$00010.}

\institutionalreview{Not applicable.}

\informedconsent{Not applicable.}

\dataavailability{An archive of the optical/IR data can be found at \url{http://j1030-field.oas.inaf.it/data.html}. ALMA data are publicly available at the ALMA archive (project ID: {2018.}1.{01601}.S). JVLA data are publicly available at the JVLA archive (project ID: VLA/18A-440).}

\acknowledgments{Q.D and I.P thank Robert Laing for the helpful suggestions about the peculiar polarization calibration performed in this work. We thank the referees for their useful suggestions. This paper makes use of the following ALMA data: {2018.} {1.}{01601.}S. ALMA is a partnership of ESO (representing its member states), NSF (USA) and NINS (Japan), together with NRC (Canada), MOST and ASIAA (Taiwan), and KASI (Republic of Korea), in cooperation with the Republic of Chile. The Joint ALMA Observatory is operated by ESO, AUI/NRAO and NAOJ.} 

\conflictsofinterest{The authors declare no conflict of~interest.} 


\end{paracol}
\reftitle{References}


\begin{thebibliography}{999}

\bibitem[{Antonuccio-Delogu} and {Silk}(2008)]{antonuccio_2008}
{Antonuccio-Delogu}, V.; {Silk}, J.
\newblock {Active galactic nuclei jet-induced feedback in galaxies---I.
  Suppression of star formation}.
\newblock {\em \mnras} {\bf 2008}, {\em 389},~1750--1762, doi:10.1111/j.1365-2966.2008.13663.x.

\bibitem[{Gaibler} \em{et~al.}(2012){Gaibler}, {Khochfar}, {Krause}, and
  {Silk}]{gaibler_2012}
{Gaibler}, V.; {Khochfar}, S.; {Krause}, M.; {Silk}, J.
\newblock {Jet-induced star formation in gas-rich galaxies}.
\newblock {\em \mnras} {\bf 2012}, {\em 425},~438--449, doi:10.1111/j.1365-2966.2012.21479.x.

\bibitem[{Salom{\'e}} \em{et~al.}(2016){Salom{\'e}}, {Salom{\'e}}, {Combes},
  and {Hamer}]{salome_2016}
{Salom{\'e}}, Q.; {Salom{\'e}}, P.; {Combes}, F.; {Hamer}, S.
\newblock {Atomic-to-molecular gas phase transition triggered by the radio jet
  in Centaurus A}.
\newblock {\em A \& A} {\bf 2016}, {\em 595},~A65, doi:10.1051/0004-6361/201628970.

\bibitem[{Gruppioni} \em{et~al.}(2011){Gruppioni}, {Pozzi}, {Zamorani}, and
  {Vignali}]{gruppioni_2011}
{Gruppioni}, C.; {Pozzi}, F.; {Zamorani}, G.; {Vignali}, C.
\newblock {Modelling galaxy and AGN evolution in the infrared: Black hole
  accretion versus star formation activity}.
\newblock {\em MNRAS} {\bf 2011}, {\em 416},~70--86, doi:10.1111/j.1365-2966.2011.19006.x.

\bibitem[{Delvecchio} \em{et~al.}(2014){Delvecchio}, {Gruppioni}, {Pozzi},
  {Berta}, {Zamorani}, {Cimatti}, {Lutz}, {Scott}, {Vignali}, {Cresci},
  {Feltre}, {Cooray}, {Vaccari}, {Fritz}, {Le Floc'h}, {Magnelli}, {Popesso},
  {Oliver}, {Bock}, {Carollo}, {Contini}, {Le F{\'e}vre}, {Lilly}, {Mainieri},
  {Renzini}, and {Scodeggio}]{delvecchio_2014}
{Delvecchio}, I.; {Gruppioni}, C.; {Pozzi}, F.; {Berta}, S.; {Zamorani}, G.;
  {Cimatti}, A.; {Lutz}, D.; {Scott}, D.; {Vignali}, C.; {Cresci}, G.; et al.
\newblock {Tracing the cosmic growth of supermassive black holes to z
  {\ensuremath{\sim}} 3 with Herschel}.
\newblock {\em \mnras} {\bf 2014}, {\em 439},~2736--2754, doi:10.1093/mnras/stu130.

\bibitem[{Rodighiero} \em{et~al.}(2015){Rodighiero}, {Brusa}, {Daddi},
  {Negrello}, {Mullaney}, {Delvecchio}, {Lutz}, {Renzini}, {Franceschini},
  {Baronchelli}, {Pozzi}, {Gruppioni}, {Strazzullo}, {Cimatti}, and
  {Silverman}]{rodighiero_2015}
{Rodighiero}, G.; {Brusa}, M.; {Daddi}, E.; {Negrello}, M.; {Mullaney}, J.R.;
  {Delvecchio}, I.; {Lutz}, D.; {Renzini}, A.; {Franceschini}, A.;
  {Baronchelli}, I.; et al.
\newblock {Relationship between Star Formation Rate and Black Hole Accretion At
  Z = 2: The Different Contributions in Quiescent, Normal, and Starburst
  Galaxies}.
\newblock {\em \apjl} {\bf 2015}, {\em 800},~L10, doi:10.1088/2041-8205/800/1/L10.

\bibitem[{Venemans} \em{et~al.}(2007){Venemans}, {R{\"o}ttgering}, {Miley},
  {van Breugel}, {de Breuck}, {Kurk}, {Pentericci}, {Stanford}, {Overzier},
  {Croft}, and {Ford}]{venemans_2007}
{Venemans}, B.P.; {R{\"o}ttgering}, H.J.A.; {Miley}, G.K.; {van Breugel},
  W.J.M.; {de Breuck}, C.; {Kurk}, J.D.; {Pentericci}, L.; {Stanford}, S.A.;
  {Overzier}, R.A.; {Croft}, S.; et al.
\newblock {Protoclusters associated with z > 2 radio galaxies . I.
  Characteristics of high redshift protoclusters}.
\newblock {\em \aap} {\bf 2007}, {\em 461},~823--845, doi:10.1051/0004-6361:20053941.

\bibitem[{Miley} and {De Breuck}(2008)]{miley_2008}
{Miley}, G.; {De Breuck}, C.
\newblock {Distant radio galaxies and their environments}.
\newblock {\em \aapr} {\bf 2008}, {\em 15},~67--144, doi:10.1007/s00159-007-0008-z.

\bibitem[{Chiaberge} \em{et~al.}(2010){Chiaberge}, {Capetti}, {Macchetto},
  {Rosati}, {Tozzi}, and {Tremblay}]{chiaberge_2010}
{Chiaberge}, M.; {Capetti}, A.; {Macchetto}, F.D.; {Rosati}, P.; {Tozzi}, P.;
  {Tremblay}, G.R.
\newblock {Three Candidate Clusters of Galaxies at Redshift
  \raisebox{-0.5ex}\textasciitilde1.8: The ``Missing Link'' Between
  Protoclusters and Local Clusters?}
\newblock {\em \apjl} {\bf 2010}, {\em 710},~L107--L110, doi:10.1088/2041-8205/710/2/L107.

\bibitem[{Fragile} \em{et~al.}(2017){Fragile}, {Anninos}, {Croft}, {Lacy}, and
  {Witry}]{fragile_2017}
{Fragile}, P.C.; {Anninos}, P.; {Croft}, S.; {Lacy}, M.; {Witry}, J.W.L.
\newblock {Numerical Simulations of a Jet-Cloud Collision and Starburst:
  Application to Minkowski{\textquoteright}s Object}.
\newblock {\em \apj} {\bf 2017}, {\em 850},~171, doi:10.3847/1538-4357/aa95c6.

\bibitem[{Overzier}(2016)]{overzier_2016}
{Overzier}, R.A.
\newblock {The realm of the galaxy protoclusters. A review}.
\newblock {\em \aapr} {\bf 2016}, {\em 24},~14, doi:10.1007/s00159-016-0100-3.

\bibitem[{Fan} \em{et~al.}(2001){Fan}, {Narayanan}, {Lupton}, {Strauss},
  {Knapp}, {Becker}, {White}, {Pentericci}, {Leggett}, {Haiman}, {Gunn},
  {Ivezi{\'c}}, {Schneider}, {Anderson}, {Brinkmann}, {Bahcall}, {Connolly},
  {Csabai}, {Doi}, {Fukugita}, {Geballe}, {Grebel}, {Harbeck}, {Hennessy},
  {Lamb}, {Miknaitis}, {Munn}, {Nichol}, {Okamura}, {Pier}, {Prada},
  {Richards}, {Szalay}, and {York}]{fan_2001}
{Fan}, X.; {Narayanan}, V.K.; {Lupton}, R.H.; {Strauss}, M.A.; {Knapp}, G.R.;
  {Becker}, R.H.; {White}, R.L.; {Pentericci}, L.; {Leggett}, S.K.; {Haiman},
  Z.; et al.
\newblock {A Survey of z > 5.8 Quasars in the Sloan Digital Sky Survey. I.
  Discovery of Three New Quasars and the Spatial Density of Luminous Quasars at
  z\raisebox{-0.5ex}\textasciitilde6}.
\newblock {\em \aj} {\bf 2001}, {\em 122},~2833--2849, doi:10.1086/324111.

\bibitem[{Fanaroff} and {Riley}(1974)]{fanaroff_1974}
{Fanaroff}, B.L.; {Riley}, J.M.
\newblock {The morphology of extragalactic radio sources of high and low
  luminosity}.
\newblock {\em \mnras} {\bf 1974}, {\em 167},~31P--36P, doi:10.1093/mnras/167.1.31P.

\bibitem[{Condon} \em{et~al.}(1998){Condon}, {Cotton}, {Greisen}, {Yin},
  {Perley}, {Taylor}, and {Broderick}]{condon_1998}
{Condon}, J.J.; {Cotton}, W.D.; {Greisen}, E.W.; {Yin}, Q.F.; {Perley}, R.A.;
  {Taylor}, G.B.; {Broderick}, J.J.
\newblock {The NRAO VLA Sky Survey}.
\newblock {\em \aj} {\bf 1998}, {\em 115},~1693--1716, doi:10.1086/300337.

\bibitem[{Gilli} \em{et~al.}(2019){Gilli}, {Mignoli}, {Peca}, {Nanni}, {Prand
  oni}, {Liuzzo}, {D'Amato}, {Brusa}, {Calura}, {Caminha}, {Chiaberge},
  {Comastri}, {Cucciati}, {Cusano}, {Grandi}, {Decarli}, {Lanzuisi},
  {Mannucci}, {Pinna}, {Tozzi}, {Vanzella}, {Vignali}, {Vito}, {Balmaverde},
  {Citro}, {Cappelluti}, {Zamorani}, and {Norman}]{gilli_2019}
{Gilli}, R.; {Mignoli}, M.; {Peca}, A.; {Nanni}, R.; {Prandoni}, I.; {Liuzzo},
  E.; {D'Amato}, Q.; {Brusa}, M.; {Calura}, F.; {Caminha}, G.B.; et al.
\newblock {Discovery of a galaxy overdensity around a powerful, heavily
  obscured FRII radio galaxy at z = 1.7: Star formation promoted by large-scale
  AGN feedback?}
\newblock {\em \aap} {\bf 2019}, {\em 632},~A26, doi:10.1051/0004-6361/201936121.

\bibitem[{D'Amato} \em{et~al.}(2020){D'Amato}, {Gilli}, {Prandoni}, {Vignali},
  {Massardi}, {Mignoli}, {Cucciati}, {Morishita}, {Decarli}, {Brusa}, {Calura},
  {Balmaverde}, {Chiaberge}, {Liuzzo}, {Nanni}, {Peca}, {Pensabene}, {Tozzi},
  and {Norman}]{damato_2020b}
{D'Amato}, Q.; {Gilli}, R.; {Prandoni}, I.; {Vignali}, C.; {Massardi}, M.;
  {Mignoli}, M.; {Cucciati}, O.; {Morishita}, T.; {Decarli}, R.; {Brusa}, \mbox{M.;
  et al}.
\newblock {Discovery of molecular gas fueling galaxy growth in a protocluster
  at z = 1.7}.
\newblock {\em \aap} {\bf 2020}, {\em 641},~L6, doi:10.1051/0004-6361/202038711.

\bibitem[{D'Amato}()]{damato_subm}
{D'Amato}, Q.; {Prandoni}, I.; {Gilli}, R.; {Vignali}, C.; {Liuzzo}, E.; {Mignoli}, M.; {Marchesi}, S.; {Peca} A.; {Chiaberge}. M; {Norman}, C; et al.
\newblock {Deep 1.4 GHz Catalogue of the J1030 Equatorial Field: A New
  Window on the Early Stages of the Universe}.
\newblock {submitted to {\em \aap}}. 
  Available online:
  \url{http://j1030-field.oas.inaf.it/publications.htm}  (accessed on 6 December 2021).

\bibitem[{de Ruiter} \em{et~al.}(1990){de Ruiter}, {Parma}, {Fanti}, and
  {Fanti}]{deruiter_1990}
{de Ruiter}, H.R.; {Parma}, P.; {Fanti}, C.; {Fanti}, R.
\newblock {VLA observations of low luminosity radio galaxies. VII---General
  properties}.
\newblock {\em \aap} {\bf 1990}, {\em 227},~351--361.

\bibitem[{Brienza}()]{brienza_subm}
{Brienza}, M.; {Gilli}, R; {Prandoni}, I.; {D'Amato}, Q.; {Vignali}, C.; {Mignoli}, M.; {Peca} A.; {Marchesi}, S.; {Liuzzo}, E.; {Norman}, C; et al.
\newblock {LOFAR Observations of the Giant FRII Radio Quasar
  J103025+052430 at the Centre of a Protocluster at z = 1.7}.
\newblock {submitted to {\em \aap}}.

\bibitem[{Planck Collaboration} \em{et~al.}(2016){Planck Collaboration}, {Ade},
  {Aghanim}, {Arnaud}, {Ashdown}, {Aumont}, {Baccigalupi}, {Banday},
  {Barreiro}, {Bartlett}, {Bartolo}, {Battaner}, {Battye}, {Benabed},
  {Beno{\^\i}t}, {Benoit-L{\'e}vy}, {Bernard}, {Bersanelli}, {Bielewicz},
  {Bock}, {Bonaldi}, {Bonavera}, {Bond}, {Borrill}, {Bouchet}, {Boulanger},
  {Bucher}, {Burigana}, {Butler}, {Calabrese}, {Cardoso}, {Catalano},
  {Challinor}, {Chamballu}, {Chary}, {Chiang}, {Chluba}, {Christensen},
  {Church}, {Clements}, {Colombi}, {Colombo}, {Combet}, {Coulais}, {Crill},
  {Curto}, {Cuttaia}, {Danese}, {Davies}, {Davis}, {de Bernardis}, {de Rosa},
  {de Zotti}, {Delabrouille}, {D{\'e}sert}, {Di Valentino}, {Dickinson},
  {Diego}, {Dolag}, {Dole}, {Donzelli}, {Dor{\'e}}, {Douspis}, {Ducout},
  {Dunkley}, {Dupac}, {Efstathiou}, {Elsner}, {En{\ss}lin}, {Eriksen},
  {Farhang}, {Fergusson}, {Finelli}, {Forni}, {Frailis}, {Fraisse},
  {Franceschi}, {Frejsel}, {Galeotta}, {Galli}, {Ganga}, {Gauthier}, {Gerbino},
  {Ghosh}, {Giard}, {Giraud-H{\'e}raud}, {Giusarma}, {Gjerl{\o}w},
  {Gonz{\'a}lez-Nuevo}, {G{\'o}rski}, {Gratton}, {Gregorio}, {Gruppuso},
  {Gudmundsson}, {Hamann}, {Hansen}, {Hanson}, {Harrison}, {Helou},
  {Henrot-Versill{\'e}}, {Hern{\'a}ndez-Monteagudo}, {Herranz}, {Hildebrand t},
  {Hivon}, {Hobson}, {Holmes}, {Hornstrup}, {Hovest}, {Huang}, {Huffenberger},
  {Hurier}, {Jaffe}, {Jaffe}, {Jones}, {Juvela}, {Keih{\"a}nen}, {Keskitalo},
  {Kisner}, {Kneissl}, {Knoche}, {Knox}, {Kunz}, {Kurki-Suonio}, {Lagache},
  {L{\"a}hteenm{\"a}ki}, {Lamarre}, {Lasenby}, {Lattanzi}, {Lawrence}, {Leahy},
  {Leonardi}, {Lesgourgues}, {Levrier}, {Lewis}, {Liguori}, {Lilje},
  {Linden-V{\o}rnle}, {L{\'o}pez-Caniego}, {Lubin}, {Mac{\'\i}as-P{\'e}rez},
  {Maggio}, {Maino}, {Mandolesi}, {Mangilli}, {Marchini}, {Maris}, {Martin},
  {Martinelli}, {Mart{\'\i}nez-Gonz{\'a}lez}, {Masi}, {Matarrese}, {McGehee},
  {Meinhold}, {Melchiorri}, {Melin}, {Mendes}, {Mennella}, {Migliaccio},
  {Millea}, {Mitra}, {Miville-Desch{\^e}nes}, {Moneti}, {Montier}, {Morgante},
  {Mortlock}, {Moss}, {Munshi}, {Murphy}, {Naselsky}, {Nati}, {Natoli},
  {Netterfield}, {N{\o}rgaard-Nielsen}, {Noviello}, {Novikov}, {Novikov},
  {Oxborrow}, {Paci}, {Pagano}, {Pajot}, {Paladini}, {Paoletti}, {Partridge},
  {Pasian}, {Patanchon}, {Pearson}, {Perdereau}, {Perotto}, {Perrotta},
  {Pettorino}, {Piacentini}, {Piat}, {Pierpaoli}, {Pietrobon}, {Plaszczynski},
  {Pointecouteau}, {Polenta}, {Popa}, {Pratt}, {Pr{\'e}zeau}, {Prunet},
  {Puget}, {Rachen}, {Reach}, {Rebolo}, {Reinecke}, {Remazeilles}, {Renault},
  {Renzi}, {Ristorcelli}, {Rocha}, {Rosset}, {Rossetti}, {Roudier},
  {Rouill{\'e} d'Orfeuil}, {Rowan-Robinson}, {Rubi{\~n}o-Mart{\'\i}n},
  {Rusholme}, {Said}, {Salvatelli}, {Salvati}, {Sandri}, {Santos},
  {Savelainen}, {Savini}, {Scott}, {Seiffert}, {Serra}, {Shellard}, {Spencer},
  {Spinelli}, {Stolyarov}, {Stompor}, {Sudiwala}, {Sunyaev}, {Sutton},
  {Suur-Uski}, {Sygnet}, {Tauber}, {Terenzi}, {Toffolatti}, {Tomasi},
  {Tristram}, {Trombetti}, {Tucci}, {Tuovinen}, {T{\"u}rler}, {Umana},
  {Valenziano}, {Valiviita}, {Van Tent}, {Vielva}, {Villa}, {Wade}, {Wandelt},
  {Wehus}, {White}, {White}, {Wilkinson}, {Yvon}, {Zacchei}, and
  {Zonca}]{PLANCK_2016}
Collaboration, P.; {Ade}, P.A.R.; {Aghanim}, N.; {Arnaud}, M.; {Ashdown},
  M.; {Aumont}, J.; {Baccigalupi}, C.; {Banday}, A.J.; {Barreiro}, R.B.;
  {Bartlett}, J.G.; et al.
\newblock {Planck 2015 results. XIII. Cosmological parameters}.
\newblock {\em \aap} {\bf 2016}, {\em 594},~A13, doi:10.1051/0004-6361/201525830.

\bibitem[Gawiser \em{et~al.}(2006)Gawiser, van Dokkum, Herrera, Maza,
  Castander, Infante, Lira, Quadri, Toner, Treister, Urry, Altmann, Assef,
  Christlein, Coppi, Duran, Franx, Galaz, Huerta, Liu, Lopez, Mendez, Moore,
  Rubio, Ruiz, Toft, and Yi]{gawiser_2006}
Gawiser, E.; van Dokkum, P.G.; Herrera, D.; Maza, J.; Castander, F.J.; Infante,
  L.; Lira, P.; Quadri, R.; Toner, R.; Treister, E.; et al.
\newblock The Multiwavelength Survey by Yale-Chile ({MUSYC}): Survey Design and
  Deep Public {UBVRI} z $\prime$ Images and Catalogs of the Extended Hubble
  Deep Field{\textendash}South.
\newblock {\em  Astrophys. J. Suppl. Ser.} {\bf 2006}, {\em
  162},~1--19, doi:10.1086/497644.

\bibitem[{Blanc} \em{et~al.}(2008){Blanc}, {Lira}, {Barrientos}, {Aguirre},
  {Francke}, {Taylor}, {Quadri}, {Marchesini}, {Infante}, {Gawiser}, {Hall},
  {Willis}, {Herrera}, {Maza}, and {MUSYC Collaboration}]{blanc_2008}
{Blanc}, G.A.; {Lira}, P.; {Barrientos}, L.F.; {Aguirre}, P.; {Francke}, H.;
  {Taylor}, E.N.; {Quadri}, R.; {Marchesini}, D.; {Infante}, L.; {Gawiser}, E.; et~al.
\newblock {The Multiwavelength Survey by Yale-Chile (MUSYC): Wide K-Band
  Imaging, Photometric Catalogs, Clustering, and Physical Properties of
  Galaxies at z \raisebox{-0.5ex}\textasciitilde 2}.
\newblock {\em \apj} {\bf 2008}, {\em 681},~1099--1115, doi:10.1086/588018.

\bibitem[{Balmaverde} \em{et~al.}(2017){Balmaverde}, {Gilli}, {Mignoli},
  {Bolzonella}, {Brusa}, {Cappelluti}, {Comastri}, {Sani}, {Vanzella},
  {Vignali}, {Vito}, and {Zamorani}]{balmaverde_2017}
{Balmaverde}, B.; {Gilli}, R.; {Mignoli}, M.; {Bolzonella}, M.; {Brusa}, M.;
  {Cappelluti}, N.; {Comastri}, A.; {Sani}, E.; {Vanzella}, E.; {Vignali}, C.; et~al.
\newblock {Primordial environment of supermassive black holes. II. Deep Y- and
  J-band images around the z 6.3 quasar SDSS J1030+0524}.
\newblock {\em \aap} {\bf 2017}, {\em 606},~A23, doi:10.1051/0004-6361/201730683.

\bibitem[{Annunziatella} \em{et~al.}(2018){Annunziatella}, {Marchesini},
  {Stefanon}, {Muzzin}, {Lange-Vagle}, {Cybulski}, {Labbe}, {Kado-Fong},
  {Bezanson}, {Brammer}, {Herrera}, {Lundgren}, {Marsan}, {Nonino}, {Rudnick},
  {Saracco}, {Tomer}, {Valdes}, {van der Burg}, {van Dokkum}, {Wake}, and
  {Whitaker}]{annunziatella_2018}
{Annunziatella}, M.; {Marchesini}, D.; {Stefanon}, M.; {Muzzin}, A.;
  {Lange-Vagle}, D.; {Cybulski}, R.; {Labbe}, I.; {Kado-Fong}, E.; {Bezanson},
  R.; {Brammer}, G.; et al.
\newblock {Complete IRAC Mapping of the CFHTLS-DEEP, MUSYC, and NMBS-II
  Fields}.
\newblock {\em \pasp} {\bf 2018}, {\em 130},~124501, doi:10.1088/1538-3873/aae796.

\bibitem[{Leipski} \em{et~al.}(2014){Leipski}, {Meisenheimer}, {Walter},
  {Klaas}, {Dannerbauer}, {De Rosa}, {Fan}, {Haas}, {Krause}, and
  {Rix}]{leipski_2014}
{Leipski}, C.; {Meisenheimer}, K.; {Walter}, F.; {Klaas}, U.; {Dannerbauer},
  H.; {De Rosa}, G.; {Fan}, X.; {Haas}, M.; {Krause}, O.; {Rix}, H.W.
\newblock {Spectral Energy Distributions of QSOs at z > 5: Common Active
  Galactic Nucleus-heated Dust and Occasionally Strong Star-formation}.
\newblock {\em \apj} {\bf 2014}, {\em 785},~154, doi:10.1088/0004-637X/785/2/154.

\bibitem[{Stiavelli} \em{et~al.}(2005){Stiavelli}, {Djorgovski}, {Pavlovsky},
  {Scarlata}, {Stern}, {Mahabal}, {Thompson}, {Dickinson}, {Panagia}, and
  {Meylan}]{stiavelli_2005}
{Stiavelli}, M.; {Djorgovski}, S.G.; {Pavlovsky}, C.; {Scarlata}, C.; {Stern},
  D.; {Mahabal}, A.; {Thompson}, D.; {Dickinson}, M.; {Panagia}, N.; {\mbox{Meylan}},
  G.
\newblock {Evidence of Primordial Clustering around the QSO SDSS J1030+0524 at
  z = 6.28}.
\newblock {\em \apjl} {\bf 2005}, {\em 622},~L1--L4, doi:10.1086/429406.

\bibitem[{Quadri} \em{et~al.}(2007){Quadri}, {Marchesini}, {van Dokkum},
  {Gawiser}, {Franx}, {Lira}, {Rudnick}, {Urry}, {Maza}, {Kriek}, {Barrientos},
  {Blanc}, {Castander}, {Christlein}, {Coppi}, {Hall}, {Herrera}, {Infante},
  {Taylor}, {Treister}, and {Willis}]{quadri_2007}
{Quadri}, R.; {Marchesini}, D.; {van Dokkum}, P.; {Gawiser}, E.; {Franx}, M.;
  {Lira}, P.; {Rudnick}, G.; {Urry}, C.M.; {Maza}, J.; {Kriek}, M.;
  et al.
\newblock {The Multiwavelength Survey by Yale-Chile (MUSYC): Deep Near-Infrared
  Imaging and the Selection of Distant Galaxies}.
\newblock {\em \aj} {\bf 2007}, {\em 134},~1103--1117, doi:10.1086/520330.

\bibitem[{Zeballos} \em{et~al.}(2018){Zeballos}, {Aretxaga}, {Hughes},
  {Humphrey}, {Wilson}, {Austermann}, {Dunlop}, {Ezawa}, {Ferrusca},
  {Hatsukade}, {Ivison}, {Kawabe}, {Kim}, {Kodama}, {Kohno}, {Monta{\~n}a},
  {Nakanishi}, {Plionis}, {S{\'a}nchez-Arg{\"u}elles}, {Stevens}, {Tamura},
  {Velazquez}, and {Yun}]{zeballos_2018}
{Zeballos}, M.; {Aretxaga}, I.; {Hughes}, D.H.; {Humphrey}, A.; {Wilson}, G.W.;
  {Austermann}, J.; {Dunlop}, J.S.; {Ezawa}, H.; {Ferrusca}, D.; {Hatsukade},
  B.; et al.
\newblock {AzTEC 1.1 mm observations of high-z protocluster environments: SMG
  overdensities and misalignment between AGN jets and SMG distribution}.
\newblock {\em \mnras} {\bf 2018}, {\em 479},~4577--4632, doi:10.1093/mnras/sty1714.

\bibitem[{McMullin} \em{et~al.}(2007){McMullin}, {Waters}, {Schiebel}, {Young},
  and {Golap}]{mcmullin_2007}
{McMullin}, J.P.; {Waters}, B.; {Schiebel}, D.; {Young}, W.; {Golap}, K.
\newblock {CASA architecture and applications}.
\newblock  In \emph{Astronomical Data Analysis Software and Systems XVI}; {Shaw}, R.A.,
  {Hill}, F., {Bell}, D.J., Eds.; Astronomical Society of
  the Pacific Conference Series; San Francisco, CA \textbf{2007}; Volume~376, p. 127.

\bibitem[{D'Amato} \em{et~al.}(2020){D'Amato}, {Gilli}, {Vignali}, {Massardi},
  {Pozzi}, {Zamorani}, {Circosta}, {Vito}, {Fritz}, {Cresci}, {Casasola},
  {Calura}, {Feltre}, {Manieri}, {Rigopoulou}, {Tozzi}, and
  {Norman}]{damato_2020}
{D'Amato}, Q.; {Gilli}, R.; {Vignali}, C.; {Massardi}, M.; {Pozzi}, F.;
  {Zamorani}, G.; {Circosta}, C.; {Vito}, F.; {Fritz}, J.; {Cresci}, G.;
  et al.
\newblock {Dust and gas content of high-redshift galaxies hosting obscured AGN
  in the CDF-S}.
\newblock {\em \aap} {\bf 2020}, {\em 636},~A37.

\bibitem[{Taylor} \em{et~al.}(2009){Taylor}, {Stil}, and
  {Sunstrum}]{taylor_2009}
{Taylor}, A.R.; {Stil}, J.M.; {Sunstrum}, C.
\newblock {A Rotation Measure Image of the Sky}.
\newblock {\em \apj} {\bf 2009}, {\em 702},~1230--1236, doi:10.1088/0004-637X/702/2/1230.

\bibitem[{George} \em{et~al.}(2012){George}, {Stil}, and {Keller}]{george_2012}
{George}, S.J.; {Stil}, J.M.; {Keller}, B.W.
\newblock {Detection Thresholds and Bias Correction in Polarized Intensity}.
\newblock {\em \pasa} {\bf 2012}, {\em 29},~214--220, doi:10.1071/AS11027.

\bibitem[{Fritz} \em{et~al.}(2006){Fritz}, {Franceschini}, and
  {Hatziminaoglou}]{fritz_2006}
{Fritz}, J.; {Franceschini}, A.; {Hatziminaoglou}, E.
\newblock {Revisiting the infrared spectra of active galactic nuclei with a new
  torus emission model}.
\newblock {\em \mnras} {\bf 2006}, {\em 366},~767--786, doi:10.1111/j.1365-2966.2006.09866.x.

\bibitem[{Feltre} \em{et~al.}(2012){Feltre}, {Hatziminaoglou}, {Fritz}, and
  {Franceschini}]{feltre_2012}
{Feltre}, A.; {Hatziminaoglou}, E.; {Fritz}, J.; {Franceschini}, A.
\newblock {Smooth and clumpy dust distributions in AGN: A direct comparison of
  two commonly explored infrared emission models}.
\newblock {\em \mnras} {\bf 2012}, {\em 426},~120--127, doi:10.1111/j.1365-2966.2012.21695.x.

\bibitem[{Circosta} \em{et~al.}(2019){Circosta}, {Vignali}, {Gilli}, {Feltre},
  {Vito}, {Calura}, {Mainieri}, {Massardi}, and {Norman}]{circosta_2019}
{Circosta}, C.; {Vignali}, C.; {Gilli}, R.; {Feltre}, A.; {Vito}, F.; {Calura},
  F.; {Mainieri}, V.; {Massardi}, M.; {Norman}, C.
\newblock {X-ray emission of z > 2.5 active galactic nuclei can be obscured by
  their host galaxies}.
\newblock {\em \aap} {\bf 2019}, {\em 623},~A172, doi:10.1051/0004-6361/201834426.

\bibitem[{Conley} \em{et~al.}(2011){Conley}, {Cooray}, {Vieira}, {Gonz{\'a}lez
  Solares}, {Kim}, {Aguirre}, {Amblard}, {Auld}, {Baker}, {Beelen}, {Blain},
  {Blundell}, {Bock}, {Bradford}, {Bridge}, {Brisbin}, {Burgarella},
  {Carpenter}, {Chanial}, {Chapin}, {Christopher}, {Clements}, {Cox},
  {Djorgovski}, {Dowell}, {Eales}, {Earle}, {Ellsworth-Bowers}, {Farrah},
  {Franceschini}, {Frayer}, {Fu}, {Gavazzi}, {Glenn}, {Griffin}, {Gurwell},
  {Halpern}, {Ibar}, {Ivison}, {Jarvis}, {Kamenetzky}, {Krips}, {Levenson},
  {Lupu}, {Mahabal}, {Maloney}, {Maraston}, {Marchetti}, {Marsden},
  {Matsuhara}, {Mortier}, {Murphy}, {Naylor}, {Neri}, {Nguyen}, {Oliver},
  {Omont}, {Page}, {Papageorgiou}, {Pearson}, {P{\'e}rez-Fournon}, {Pohlen},
  {Rangwala}, {Rawlings}, {Raymond}, {Riechers}, {Rodighiero}, {Roseboom},
  {Rowan-Robinson}, {Schulz}, {Scott}, {Scott}, {Serra}, {Seymour}, {Shupe},
  {Smith}, {Symeonidis}, {Tugwell}, {Vaccari}, {Valiante}, {Valtchanov},
  {Verma}, {Viero}, {Vigroux}, {Wang}, {Wiebe}, {Wright}, {Xu}, {Zeimann},
  {Zemcov}, and {Zmuidzinas}]{conley_2011}
{Conley}, A.; {Cooray}, A.; {Vieira}, J.D.; {Gonz{\'a}lez Solares}, E.A.;
  {Kim}, S.; {Aguirre}, J.E.; {Amblard}, A.; {Auld}, R.; {Baker}, A.J.;
  {Beelen}, A.; et al.
\newblock {Discovery of a Multiply Lensed Submillimeter Galaxy in Early HerMES
  Herschel/SPIRE Data}.
\newblock {\em ApJl} {\bf 2011}, {\em 732},~L35, doi:10.1088/2041-8205/732/2/L35.

\bibitem[{Fu} \em{et~al.}(2012){Fu}, {Jullo}, {Cooray}, {Bussmann}, {Ivison},
  {P{\'e}rez-Fournon}, {Djorgovski}, {Scoville}, {Yan}, {Riechers}, {Aguirre},
  {Auld}, {Baes}, {Baker}, {Bradford}, {Cava}, {Clements}, {Dannerbauer},
  {Dariush}, {De Zotti}, {Dole}, {Dunne}, {Dye}, {Eales}, {Frayer}, {Gavazzi},
  {Gurwell}, {Harris}, {Herranz}, {Hopwood}, {Hoyos}, {Ibar}, {Jarvis}, {Kim},
  {Leeuw}, {Lupu}, {Maddox}, {Mart{\'\i}nez-Navajas}, {Micha{\l}owski},
  {Negrello}, {Omont}, {Rosenman}, {Scott}, {Serjeant}, {Smail}, {Swinbank},
  {Valiante}, {Verma}, {Vieira}, {Wardlow}, and {van der Werf}]{fu_2012}
{Fu}, H.; {Jullo}, E.; {Cooray}, A.; {Bussmann}, R.S.; {Ivison}, R.J.;
  {P{\'e}rez-Fournon}, I.; {Djorgovski}, S.G.; {Scoville}, N.; {Yan}, L.;
  {Riechers}, D.A.; et al.
\newblock {A Comprehensive View of a Strongly Lensed Planck-Associated
  Submillimeter Galaxy}.
\newblock {\em \apj} {\bf 2012}, {\em 753},~134, doi:10.1088/0004-637X/753/2/134.

\bibitem[{Riechers} \em{et~al.}(2013){Riechers}, {Bradford}, {Clements},
  {Dowell}, {P{\'e}rez-Fournon}, {Ivison}, {Bridge}, {Conley}, {Fu}, {Vieira},
  {Wardlow}, {Calanog}, {Cooray}, {Hurley}, {Neri}, {Kamenetzky}, {Aguirre},
  {Altieri}, {Arumugam}, {Benford}, {B{\'e}thermin}, {Bock}, {Burgarella},
  {Cabrera-Lavers}, {Chapman}, {Cox}, {Dunlop}, {Earle}, {Farrah}, {Ferrero},
  {Franceschini}, {Gavazzi}, {Glenn}, {Solares}, {Gurwell}, {Halpern},
  {Hatziminaoglou}, {Hyde}, {Ibar}, {Kov{\'a}cs}, {Krips}, {Lupu}, {Maloney},
  {Martinez-Navajas}, {Matsuhara}, {Murphy}, {Naylor}, {Nguyen}, {Oliver},
  {Omont}, {Page}, {Petitpas}, {Rangwala}, {Roseboom}, {Scott}, {Smith},
  {Staguhn}, {Streblyanska}, {Thomson}, {Valtchanov}, {Viero}, {Wang},
  {Zemcov}, and {Zmuidzinas}]{riechers_2013}
{Riechers}, D.A.; {Bradford}, C.M.; {Clements}, D.L.; {Dowell}, C.D.;
  {P{\'e}rez-Fournon}, I.; {Ivison}, R.J.; {Bridge}, C.; {Conley}, A.; {Fu},
  H.; {Vieira}, J.D.; et al.
\newblock {A dust-obscured massive maximum-starburst galaxy at a redshift of
  6.34}.
\newblock {\em \nat} {\bf 2013}, {\em 496},~329--333, doi:10.1038/nature12050.

\bibitem[{Gilli} \em{et~al.}(2014){Gilli}, {Norman}, {Vignali}, {Vanzella},
  {Calura}, {Pozzi}, {Massardi}, {Mignano}, {Casasola}, {Daddi}, {Elbaz},
  {Dickinson}, {Iwasawa}, {Maiolino}, {Brusa}, {Vito}, {Fritz}, {Feltre},
  {Cresci}, {Mignoli}, {Comastri}, and {Zamorani}]{gilli_2014}
{Gilli}, R.; {Norman}, C.; {Vignali}, C.; {Vanzella}, E.; {Calura}, F.;
  {Pozzi}, F.; {Massardi}, M.; {Mignano}, A.; {Casasola}, V.; {Daddi}, E.;
  et al.
\newblock {ALMA reveals a warm and compact starburst around a heavily obscured
  supermassive black hole at z = 4.75}.
\newblock {\em A\&A} {\bf 2014}, {\em 562},~A67, doi:10.1051/0004-6361/201322892.

\bibitem[{Rangwala} \em{et~al.}(2011){Rangwala}, {Maloney}, {Glenn}, {Wilson},
  {Rykala}, {Isaak}, {Baes}, {Bendo}, {Boselli}, {Bradford}, {Clements},
  {Cooray}, {Fulton}, {Imhof}, {Kamenetzky}, {Madden}, {Mentuch}, {Sacchi},
  {Sauvage}, {Schirm}, {Smith}, {Spinoglio}, and {Wolfire}]{rangwala_2011}
{Rangwala}, N.; {Maloney}, P.R.; {Glenn}, J.; {Wilson}, C.D.; {Rykala}, A.;
  {Isaak}, K.; {Baes}, M.; {Bendo}, G.J.; {Boselli}, A.; {Bradford}, C.M.;
  et al.
\newblock {Observations of Arp 220 Using Herschel-SPIRE: An Unprecedented View
  of the Molecular Gas in an Extreme Star Formation Environment}.
\newblock {\em ApJ} {\bf 2011}, {\em 743},~94, doi:10.1088/0004-637X/743/1/94.

\bibitem[{Padovani} \em{et~al.}(2017){Padovani}, {Alexander}, {Assef}, {De
  Marco}, {Giommi}, {Hickox}, {Richards}, {Smol{\v c}i{\'c}}, {Hatziminaoglou},
  {Mainieri}, and {Salvato}]{padovani_2017}
{Padovani}, P.; {Alexander}, D.M.; {Assef}, R.J.; {De Marco}, B.; {Giommi}, P.;
  {Hickox}, R.C.; {Richards}, G.T.; {Smol{\v c}i{\'c}}, V.; {Hatziminaoglou},
  E.; {Mainieri}, V.; et al.
\newblock {Active galactic nuclei: What's in a name?}
\newblock {\em A\&AR} {\bf 2017}, {\em 25},~2, doi:10.1007/s00159-017-0102-9.

\bibitem[{Kennicutt}(1998)]{kennicutt_1998}
{Kennicutt}, R.C., Jr.
\newblock {The Global Schmidt Law in Star-forming Galaxies}.
\newblock {\em \apj} {\bf 1998}, {\em 498},~541--552, doi:10.1086/305588.

\bibitem[{Duras} \em{et~al.}(2020){Duras}, {Bongiorno}, {Ricci}, {Piconcelli},
  {Shankar}, {Lusso}, {Bianchi}, {Fiore}, {Maiolino}, {Marconi}, {Onori},
  {Sani}, {Schneider}, {Vignali}, and {La Franca}]{duras_2020}
{Duras}, F.; {Bongiorno}, A.; {Ricci}, F.; {Piconcelli}, E.; {Shankar}, F.;
  {Lusso}, E.; {Bianchi}, S.; {Fiore}, F.; {Maiolino}, R.; {Marconi}, A.;
  et al.
\newblock {Universal bolometric corrections for active galactic nuclei over
  seven luminosity decades}.
\newblock {\em \aap} {\bf 2020}, {\em 636},~A73, doi:10.1051/0004-6361/201936817.

\bibitem[{Lusso} \em{et~al.}(2012){Lusso}, {Comastri}, {Simmons}, {Mignoli},
  {Zamorani}, {Vignali}, {Brusa}, {Shankar}, {Lutz}, {Trump}, {Maiolino},
  {Gilli}, {Bolzonella}, {Puccetti}, {Salvato}, {Impey}, {Civano}, {Elvis},
  {Mainieri}, {Silverman}, {Koekemoer}, {Bongiorno}, {Merloni}, {Berta}, {Le
  Floc'h}, {Magnelli}, {Pozzi}, and {Riguccini}]{lusso_2012}
{Lusso}, E.; {Comastri}, A.; {Simmons}, B.D.; {Mignoli}, M.; {Zamorani}, G.;
  {Vignali}, C.; {Brusa}, M.; {Shankar}, F.; {Lutz}, D.; {Trump}, J.R.;
  et al.
\newblock {Bolometric luminosities and Eddington ratios of X-ray selected
  active galactic nuclei in the XMM-COSMOS survey}.
\newblock {\em \mnras} {\bf 2012}, {\em 425},~623--640, doi:10.1111/j.1365-2966.2012.21513.x.

\bibitem[{Schreiber} \em{et~al.}(2015){Schreiber}, {Pannella}, {Elbaz},
  {B{\'e}thermin}, {Inami}, {Dickinson}, {Magnelli}, {Wang}, {Aussel}, {Daddi},
  {Juneau}, {Shu}, {Sargent}, {Buat}, {Faber}, {Ferguson}, {Giavalisco},
  {Koekemoer}, {Magdis}, {Morrison}, {Papovich}, {Santini}, and
  {Scott}]{schreiber_2015}
{Schreiber}, C.; {Pannella}, M.; {Elbaz}, D.; {B{\'e}thermin}, M.; {Inami}, H.;
  {Dickinson}, M.; {Magnelli}, B.; {Wang}, T.; {Aussel}, H.; {Daddi}, E.;
  et al.
\newblock {The Herschel view of the dominant mode of galaxy growth from z = 4
  to the present day}.
\newblock {\em \aap} {\bf 2015}, {\em 575},~A74, doi:10.1051/0004-6361/201425017.

\bibitem[{Lapi} \em{et~al.}(2018){Lapi}, {Pantoni}, {Zanisi}, {Shi}, {Mancuso},
  {Massardi}, {Shankar}, {Bressan}, and {Danese}]{lapi_2018}
{Lapi}, A.; {Pantoni}, L.; {Zanisi}, L.; {Shi}, J.; {Mancuso}, C.; {Massardi},
  M.; {Shankar}, F.; {Bressan}, A.; {Danese}, L.
\newblock {The Dramatic Size and Kinematic Evolution of Massive Early-type
  Galaxies}.
\newblock {\em ApJ} {\bf 2018}, {\em 857},~22, doi:10.3847/1538-4357/aab6af.

\bibitem[{Blandford} and {Znajek}(1977)]{blandford_1977}
{Blandford}, R.D.; {Znajek}, R.L.
\newblock {Electromagnetic extraction of energy from Kerr black holes.}
\newblock {\em \mnras} {\bf 1977}, {\em 179},~433--456, doi:10.1093/mnras/179.3.433.

\bibitem[{Ghisellini} \em{et~al.}(2014){Ghisellini}, {Tavecchio}, {Maraschi},
  {Celotti}, and {Sbarrato}]{ghisellini_2014}
{Ghisellini}, G.; {Tavecchio}, F.; {Maraschi}, L.; {Celotti}, A.; {Sbarrato},
  T.
\newblock {The power of relativistic jets is larger than the luminosity of
  their accretion disks}.
\newblock {\em \nat} {\bf 2014}, {\em 515},~376--378, doi:10.1038/nature13856.

\bibitem[{Delvecchio} \em{et~al.}(2021){Delvecchio}, {Daddi}, {Sargent},
  {Jarvis}, {Elbaz}, {Jin}, {Liu}, {Whittam}, {Algera}, {Carraro}, {D'Eugenio},
  {Delhaize}, {Kalita}, {Leslie}, {Moln{\'a}r}, {Novak}, {Prandoni},
  {Smol{\v{c}}i{\'c}}, {Ao}, {Aravena}, {Bournaud}, {Collier},
  {Randriamampandry}, {Randriamanakoto}, {Rodighiero}, {Schober}, {White}, and
  {Zamorani}]{delvecchio_2021}
{Delvecchio}, I.; {Daddi}, E.; {Sargent}, M.T.; {Jarvis}, M.J.; {Elbaz}, D.;
  {Jin}, S.; {Liu}, D.; {Whittam}, I.H.; {Algera}, H.; {Carraro}, R.;
  et al.
\newblock {The infrared-radio correlation of star-forming galaxies is strongly
  M$_{{\ensuremath{\star}}}$-dependent but nearly redshift-invariant since z
  {\ensuremath{\sim}} 4}.
\newblock {\em \aap} {\bf 2021}, {\em 647},~A123, doi:10.1051/0004-6361/202039647.

\bibitem[{Giovannini} \em{et~al.}(2001){Giovannini}, {Cotton}, {Feretti},
  {Lara}, and {Venturi}]{giovannini_2001}
{Giovannini}, G.; {Cotton}, W.D.; {Feretti}, L.; {Lara}, L.; {Venturi}, T.
\newblock {VLBI Observations of a Complete Sample of Radio Galaxies: 10~Years
  Later}.
\newblock {\em \apj} {\bf 2001}, {\em 552},~508--526, doi:10.1086/320581.

\bibitem[{Ruffa} \em{et~al.}(2019){Ruffa}, {Prandoni}, {Laing}, {Paladino},
  {Parma}, {de Ruiter}, {Mignano}, {Davis}, {Bureau}, and {Warren}]{ruffa_2019}
{Ruffa}, I.; {Prandoni}, I.; {Laing}, R.A.; {Paladino}, R.; {Parma}, P.; {de
  Ruiter}, H.; {Mignano}, A.; {Davis}, T.A.; {Bureau}, M.; {Warren}, J.
\newblock {The AGN fuelling/feedback cycle in nearby radio galaxies I. ALMA
  observations and early results}.
\newblock {\em \mnras} {\bf 2019}, {\em 484},~4239--4259, doi:10.1093/mnras/stz255.

\bibitem[{Parma} \em{et~al.}(1993){Parma}, {Morganti}, {Capetti}, {Fanti}, and
  {de Ruiter}]{parma_1993}
{Parma}, P.; {Morganti}, R.; {Capetti}, A.; {Fanti}, R.; {de Ruiter}, H.R.
\newblock {Polarization properties at 1.4 GHz of low luminosity radio
  galaxies.}
\newblock {\em \aap} {\bf 1993}, {\em 267},~31--38.

\bibitem[{Muxlow} and {Garrington}(1991)]{muxlow_1991}
{Muxlow}, T.W.B.; {Garrington}, S.T. {Observations of large scale
  extragalactic jets}.
\newblock In {\em Beams and Jets in Astrophysics}; Cambridge University Press \& Assessment, Cambridge, UK 
\textbf{1991}; Volume~19, p.~52.

\bibitem[{Pentericci} \em{et~al.}(2000){Pentericci}, {Van Reeven}, {Carilli},
  {R{\"o}ttgering}, and {Miley}]{pentericci_2000}
{Pentericci}, L.; {Van Reeven}, W.; {Carilli}, C.L.; {R{\"o}ttgering}, H.J.A.;
  {Miley}, G.K.
\newblock {VLA radio continuum observations of a new sample of high redshift
  radio galaxies}.
\newblock {\em \aaps} {\bf 2000}, {\em 145},~121--159, doi:10.1051/aas:2000104.

\bibitem[{Saikia} and {Salter}(1988)]{saikia_1988}
{Saikia}, D.J.; {Salter}, C.J.
\newblock {Polarization properties of extragalactic radio sources.}
\newblock {\em \araa} {\bf 1988}, {\em 26},~93--144.

\end{thebibliography}
\end{document}